\documentstyle[11pt,amssymb,seceq]{article}
\newcommand{\be}{\begin{equation}}
\newcommand{\ee}{\end{equation}}
\newcommand{\ba}{\begin{eqnarray}}
\newcommand{\ea}{\end{eqnarray}}

\newcommand{\Alpha}{\mbox{$A$}}

\newcommand{\lb}{\label}

\newcommand{\hr}{{\widehat{r}}}

\newcommand{\bk}{{\bf k}}

\newcommand{\bq}{{\bf q}}
\newcommand{\br}{{\bf r}}
\newcommand{\bv}{{\bf v}}

\newcommand{\bD}{{\bf D}}

\newcommand{\bW}{{\bf W}}

\newcommand{\borho}{{\mbox{\boldmath $\rho$}}}
\newcommand{\bokappa}{{\mbox{\boldmath $\kappa$}}}

\newcommand{\grad}{{\mbox{\boldmath $\nabla$}}}
\newcommand{\bdot}{{\mbox{\boldmath $\cdot$}}}
\newcommand{\bzed}{{\mbox{\boldmath $0$}}}

\author{Gregory L. Eyink \footnote{Department of Mathematics,
University of Arizona, Tucson, AZ 85721. {\it
eyink@math.arizona.edu}}\hspace{.1 in} and \hspace{.1 in}
Jack Xin \footnote{ Department of Mathematics,
University of Texas at Austin,
Austin, TX 78712. {\it jxin@fireant.ma.utexas.edu}}}
\date{ }
\begin{document}
\title{Self-Similar Decay \\ in the Kraichnan Model of a Passive Scalar}

\maketitle
\begin{abstract}
We study the two-point correlation function of a freely decaying scalar in
Kraichnan's model of advection by a Gaussian random velocity field, stationary
and white-noise in time but fractional Brownian in space with roughness
exponent
$0<\zeta<2$, appropriate to the inertial-convective range of the scalar. We
find all self-similar
solutions, by transforming the scaling equation to Kummer's equation.
It is shown that only those scaling solutions with scalar energy decay
exponent $a\leq (d/\gamma)+1$ are statistically realizable,
where $d$ is space dimension and $\gamma=2-\zeta$. An infinite sequence
of invariants $J_p,\,\,p=0,1,2,...$ is pointed out, where $J_0$ is Corrsin's
integral
invariant but the higher invariants appear to be new. We show that at
least one of the invariants $J_0$ or $J_1$ must be nonzero (possibly infinite)
for realizable initial data.
Initial data with a finite, nonzero invariant---the first being
$J_p$---converge at long
times to a scaling solution $\Phi_p$ with $a=(d/\gamma)+p,\,\,p=0,1.$ The
latter belong to
an exceptional series of self-similar solutions with stretched-exponential
decay in space.
However, the domain of attraction includes many initial data with power-law
decay. When the initial
data has all invariants zero or infinite and also it exhibits power-law decay,
then the solution
converges at long times to a non-exceptional scaling solution with the same
power-law decay.
These results support a picture of a ``two-scale'' decay with breakdown of
self-similarity
for a range of exponents $(d+\gamma)/\gamma < a < (d+2)/\gamma,$ analogous to
what has recently
been found in decay of Burgers turbulence.
\end{abstract}

\newpage

\begin{center}

{\bf TABLE OF CONTENTS}

\end{center}

\noindent{\bf 1. Introduction}

\noindent{\bf 2. Background Material}

{\em (2.1) Review of the Kraichnan
Model}.......................................................pp.08-11

{\em (2.2) Phenomenology of Turbulent
Decay}.................................................pp.12-16

\noindent{\bf 3. Self-Similar Decay and Its Breakdown}

{\em (3.1) Derivation and Solution of the Scaling
Equation}.............................pp.17-19

{\em (3.2) Asymptotic Behaviors \& Permanence of Large
Eddies}....................pp.20-25

{\em (3.3) Realizability of the Scaling
Solutions}................................................pp.26-30

{\em (3.4) A Physical Explanation of the
Results}..............................................pp.31-34

\noindent{\bf 4. Convergence to Self-Similar Solutions}

{\em (4.1) Long-Time Scaling Limit}$\!\!$
..................................................................pp.35-36

{\em (4.2) Initial Data with Rapid
Decay}..........................................................pp.36-40

{\em (4.3) Initial Data with Slow Decay \& A Finite
Invariant}.........................$\,$pp.40-52

{\em (4.4) Initial Data with Slow Decay \& No Finite
Invariant}$\,$.......................pp.53-57

{\em (4.5) View On a Larger
Length-Scale}........................................................pp.57-58

\noindent {\bf 5. Conclusions}

\noindent {\bf Acknowledgements}

\noindent{\bf Appendix: Self-Similar Scalar Spectra for a Brownian Velocity
Field}

\newpage

\section{Introduction}

A key hypothesis in the theory of turbulence decay is that the energy spectrum
has
asymptotically a {\it self-similar} form
\be E(k,t)= v^2(t)\ell(t) F(k\ell(t)) \lb{O1} \ee
where $\ell(t)$ is a suitable length-scale and $v(t)$ a velocity scale.
Equivalently,
the hypothesis may be made for the longitudinal velocity two-point correlation
$B_{LL}(r,t):= \hat{r}_i\hat{r}_j\langle v_i({\bf x})v_j({\bf x}+\br)\rangle$,
that it obey:
\be B_{LL}(r,t)= v^2(t) f(r/\ell(t)). \lb{O2} \ee
Historically, hypotheses (\ref{O1}) and (\ref{O2}) were first proposed for
freely decaying turbulence
in 1938 by von K\'{a}rm\'{a}n and Howarth \cite{vKH}. They referred to the {\it
Ans\"{a}tze}
(\ref{O1}) and (\ref{O2}) as ``self-preservation hypotheses,'' since the shape
of the spectrum
and correlation function are thereby preserved in the decay process. Detailed
discussion
of such self-preservation hypotheses is to be found in \cite{MY}, Chapter 16.
Since the early work
in turbulence, corresponding hypotheses have been proposed for many other
nonequilibrium processes,
e.g. surface growth \cite{KrSp} and phase-ordering dynamics \cite{Bray}. In
those fields the assumption
is usually called {\it dynamic self-similarity} or {\it dynamic scaling}.
Essentially, the hypothesis
amounts to the statement that there is only one relevant length-scale in the
decay process. For
example, in turbulence decay this is plausibly the integral length-scale
$L(t):= {{1}\over{B_{LL}(0,t)}}
\int_0^\infty dr\,\,B_{LL}(r,t).$ Althoughly widely employed in nonequilibrium
physics, the validity
of such self-similarity hypotheses is still debated and their foundations
poorly understood.
This is particularly true when the random initial data of the system exhibit
long-range power-law
correlations.

Recently, the validity of the self-similarity has been examined in a soluble
model, the decaying
Burgers turbulence \cite{GSAFT}. Those authors solved exactly for the two-point
correlations
and energy spectra of the one-dimensional Burgers equation with initial energy
spectra exhibiting
a low-wavenumber power-law form, $E(k,t_0)\sim A k^n$ for $kL_0\ll 1$, and thus
a power-law decay
in the spatial velocity correlation function, $B(r,t_0)\sim A' r^{-(n+1)}$ for
$r\gg L_0$
(when $n$ is not a positive, even integer). What was discovered by the authors
of \cite{GSAFT}
was that the hypothesis of dynamic self-similarity was violated when $1<n<2$.
Instead, a new
length-scale $L_*(t)$ developed dynamically which was much larger
asymptotically than the integral
length-scale $L(t)$. The new length-scale was characterized by the property
that the initial
low-wavenumber power-law spectrum was preserved only for $kL_*(t)\ll 1$. This
preservation was
traditionally believed to hold for all $kL(t)\ll 1,$ which has been called the
principle
of {\it permanence of large-eddies} \cite{Frisch}. The development of two
distinct length-scales
had important consequences for the decay process. For example, the rate of
decay was found to
be different than that predicted by traditional phenomenological theory.

Another exactly soluble turbulence model is available, a model of a turbulently
advected scalar
proposed by R. H. Kraichnan in 1968 \cite{Kr68}. This model corresponds to a
stochastic
advection-diffusion equation
\be \partial_t\theta(\br,t)+(\bv(\br,t)\bdot\grad)\theta(\br,t) =
\kappa\bigtriangleup\theta(\br,t),
                                                                     \lb{1.1}
\ee
in which the advecting field $\bv(\br,t)$ is a ``synthetic turbulence'',
specifically,
a Gaussian random velocity field with zero mean and covariance
\be \langle v_i(\br,t)v_j(\br',t')\rangle = D_{ij}(\br-\br')\delta(t-t')
\lb{1.2} \ee
which is white-noise in time. The remarkable feature of the model which
Kraichnan
discovered is that there is no closure problem. In particular, Kraichnan showed
that the
2-point correlation function
$\Theta(\br,t):=\langle\theta(\br,t)\theta(\bzed,t)\rangle$
obeys the following closed equation in homogeneous scalar decay:
\be \partial_t\Theta(\br,t) =
[D_{ij}(\bzed)-D_{ij}(\br)]\grad_{\br_i}\grad_{\br_j}
                                    \Theta(\br,t)+
2\kappa\bigtriangleup\Theta(\br,t). \lb{1.3a} \ee
Recently the study of the Kraichnan model has undergone a renaissance, impelled
by the observation
\cite{Kr94} that the $N$th-order statistical correlations for $N>2$ should
exhibit ``anomalous
scaling'', not predicted by naive dimensional analysis as in Kolmogorov's 1941
theory. In particular,
attention has been focused on the {\it inertial-convective range} of the model,
in which the
molecular diffusivity $\kappa\rightarrow 0$ and the velocity covariance in
space has a power-law form
\be D_{ij}(\br)\sim D_{ij}(\bzed)- D_1\cdot r^{\zeta}\left[\delta_{ij}
+{{\zeta}\over{d-1}}\left(\delta_{ij}-{{r_ir_j}\over{r^2}}\right)\right]
          + O\left({{r}\over{L}}\right)^2, \lb{1.4} \ee
for $0<\zeta<2$ and $r\ll L$, where the latter length-scale is the velocity
integral scale.
In this relation statistical isotropy as well as homogeneity has been assumed.
The formula
(\ref{1.4}) mimicks the situation in a real turbulent scalar decay when the
scalar is spectrally
supported on length-scales $L_\theta$ much smaller than the velocity integral
scale $L$ but yet much
larger than the dissipation length-scale $\eta_d$ set by the molecular
diffusivity $\kappa$.
In this situation, the calculation of the anomalous exponents has proved to be
possible analytically
by perturbation expansion in three regimes: small space H\"{o}lder exponent
$H={{\zeta}\over{2}}$
corresponding to a ``rough'' velocity field \cite{GK95,DGK}, small exponent
$1-H$ coresponding
to expansion about a ``smooth'' velocity field or so-called Batchelor limit
\cite{SS95,SS96},
and expansion in ${{1}\over{d}}$ with $d$ the space dimension \cite{CFKL,CF}.

Our interest here is instead to study the Kraichnan model as a soluble test
case of turbulent decay.
Passive scalars undergo a turbulent decay which is similar in many respects to
that of
the velocity itself. This is very well described in \cite{LMC}. The decrease in
the scalar
energy or intensity $E_\theta(t):=
{{1}\over{2}}\langle\theta^2(\bzed,t)\rangle$
is by a process of progressive degradation of the scalar at the higher
wavenumbers. In the
course of this decay process, the scalar integral length-scale
$L_\theta(t)={{1}\over{\Theta(0,t)}}
\int_0^\infty dr\,\,\Theta(r,t)$ grows as the spectral support of the scalar is
shifted progressively
to lower wavenumbers. On dimensional grounds, this growth of the length-scale
is governed by
\be {{1}\over{L_\theta(t)}}{{dL_\theta(t)}\over{dt}}\propto D_1
L_\theta^{-\gamma}(t), \lb{1.6} \ee
with $\gamma:=2-\zeta,$ which leads to the relation $L_\theta(t)\propto
(D_1(t-t_0))^{1/\gamma}$.
As has been emphasized in \cite{LMC}, this can be thought of as a ``Richardson
diffusion''
of the scalar integral length-scale up through the velocity inertial range
(which is here
taken to be statistically stationary). This growth law for the scalar
length-scale can be
converted into an energy decay law under two additional assumptions. First, if
one considers
initial scalar spectra of a power-law form $E_\theta(k,t_0)\propto A k^n$ for
low-wavenumbers
$kL_\theta(t_0)\ll 1$, then the hypothesis of permanence of large eddies would
imply
that this low-wavenumber spectrum persists with a time-independent constant $A$
for
$kL_{\theta}^*(t)\ll 1$, if $n<d+1.$  Second, the hypothesis of dynamic
self-similarity would
imply that there is only one relevant length-scale, the scalar integral length
$L_\theta(t),$
so that $L_{\theta}^*(t)=L_\theta(t)$ up
to a constant factor. Under these two assumptions, the scalar energy may be
estimated to order
of magnitude by integrating over the low-wavenumber range up to
$L_\theta^{-1}(t)$, with the result
\be E_\theta(t)\propto A L_\theta^{-p}(t) \propto A (D_1(t-t_0))^{-p/\gamma},
\lb{1.7} \ee
for $p=n+1$. Thus, the decay rate is non-universal, and depends upon the
low-wavenumber
spectral exponent.

The main aim of this work is to examine the decay problem in the Kraichnan
model, to investigate
the universality of the ``two-scale'' phenomenon discovered in \cite{GSAFT}.
Because the standard
phenomenology is common to both velocity and passive scalar decay, we may use
the Kraichnan model as
a source of insight. There seems to have been less work on the decay problem in
the Kraichnan model
than on the statistical steady-state and most of this in the Batchelor limit
$\zeta=2$, the so-called
viscous-convective range. In addition to the early work of Kraichnan
\cite{Kr74}, the decay
of the passive scalar in the Batchelor limit has been recently examined by Son
\cite{Son}.
Our work here will be devoted instead to the inertial-convective range of the
scalar in which
the velocity correlator is given by (\ref{1.4}) with $0<\zeta<2$. A few
preliminary investigations
on this problem have been reported in a very recent review article of Majda and
Kramer \cite{MajKram},
Section 4.2, but no exhaustive study seems yet to have been made. Our interest
is thus in the low-order
$N=2$ correlator, rather than in the higher-order statistics. In the same
inertial-convective range
considered above, with the assumptions of homogeneity and isotropy, equation
(\ref{1.3a})
for the 2-point correlator simplifies to
\be \partial_t\Theta(r,t) = {{D_1}\over{r^{d-1}}}{{\partial}\over{\partial r}}
     \left[ r^{d+\zeta-1}{{\partial\Theta}\over{\partial r}}(r,t)\right].
\lb{1.3} \ee
We shall be particularly interested in the issue of self-similarity of the
decay process. In fact,
one of our main results will be an analytical construction and complete
classification of {\it all}
self-similar decay solutions of equation (\ref{1.3}), along with an analysis of
their domains of
attraction.

The reader should note that in equation (\ref{1.3}) the limit has been taken of
vanishing
molecular diffusivity. We have done so in order to focus on the turbulent
dissipation of the scalar,
which leads to a decaying scalar energy $E_\theta(t)\rightarrow 0$ as
$t\rightarrow \infty$ even
in the limit $\kappa\rightarrow 0$. This corresponds to a famous conjecture on
the three-dimensional
energy cascade put forth by Onsager \cite{On49}, who proposed that the limiting
turbulent ensemble
in the limit of vanishing viscosity $\nu\rightarrow 0$ should consist of
realizations of the inviscid
Euler equations which dissipate energy. Of course, these must be weak or
distributional solutions,
not classical solutions. Onsager coined the term ``ideal turbulence'' for this
limiting dissipative
ensemble governed by the ideal fluid equations. More recently, this ideal
mechanism of dissipation
has been called the ``dissipative anomaly,'' since Polyakov pointed out a close
similarity
to conservation-law anomalies in quantum field theory \cite{Poly}. This
specifically turbulent
mechanism of dissipation is well-illustrated by the Kraichnan model. Since the
operator on the RHS
of (\ref{1.3}) is homogeneous degree $-\gamma$, it follows using $E_\theta(t)
={{1}\over{2}}
\Theta(0,t)$ that
\be {{dE_\theta}\over{dt}}(t) = \left. {{D_1}\over{2
r^{d-1}}}{{\partial}\over{\partial r}}
     \left[ r^{d+\zeta-1}{{\partial\Theta}\over{\partial
r}}(r,t)\right]\right|_{r=0}=0 \lb{1.10} \ee
when the 2nd-order structure function
\begin{eqnarray}
S_2(r,t) & := & \langle [\theta(\br,t)-\theta(\bzed,t)]^2\rangle \cr
     \,&  = & 2[\Theta(0,t)-\Theta(r,t)]\sim C r^\xi, \lb{1.11}
\end{eqnarray}
with $\xi>\gamma$. Thus, some critical degree of singularity is required for
turbulent dissipation.
Our self-similar decay solutions---which are exact solutions of the
zero-diffusion Kraichnan equations
---explicitly illustrate this ideal dissipation mechanism. The implications for
the theory
of weak solutions of the hyperbolic stochastic PDE, the $\kappa\rightarrow 0$
limit of (\ref{1.1}),
will be discussed in a forthcoming work \cite{EX2}.

\newpage

\section{Background Material}

\noindent {\it (2.1) Review of the Kraichnan Model}

\noindent The model we consider is the stochastic partial differential equation
\be d\theta(\br,t) =
\kappa\bigtriangleup\theta(\br,t)dt-(\bW(\br,dt)\bdot\grad)\theta(\br,t) ,
                                                                   \lb{2.1} \ee
where $\bW(\cdot,t)$ is a Wiener process in the function space $C({\Bbb
R}^d,{\Bbb R}^d)$,
with covariance function
\be \langle W_i(\br,t)W_j(\br',t')\rangle =D_{ij}(\br-\br')t\wedge t' .
\lb{Vcov} \ee
The stochastic PDE is interpreted in the Stratonovich sense. See \cite{Kunita},
Chapter 6, and \cite{LJR}
for a more detailed discussion of the mathematical foundations. The spatial
covariance matrix $\bD$
we consider is defined by the Fourier integral
\be D_{ij}(\br)= D\int d^d\bk
\,\,{{P_{ij}(\bk)}\over{\left(k^2+k_0^2\right)^{(d+\zeta)/2}}}
    e^{i\bk\bdot\br}. \lb{Vspec} \ee
where $0<\zeta<2$. The constant $k_0$ is an infrared cutoff for the velocity
field,
proportional to the inverse velocity integral length $k_0\propto L^{-1}$.
$P_{ij}(\bk)$ is
the projection in ${\Bbb R}^d$ onto the subspace perpendicular to $\bk$. Thus
(\ref{Vspec})
automatically defines a suitable positive-definite, symmetric matrix-valued
function,
divergence-free in each index. We have made the choice (\ref{Vspec}) just for
specificity.
In fact, any velocity covariance with the properties discussed next would
suffice.

The matrix $D_{ij}(\br)$ can be written as
$D_{ij}(\br)=P_{ij}(\grad_\br)K(\br)$, or as
\be D_{ij}(\br)= K(r)\delta_{ij}+\partial_i\partial_jH(r), \lb{Vrep} \ee
where the function $K(r)$ is defined by the integral
\be K(r)= D\int d^d\bk\,\,{{
e^{i\bk\bdot\br}}\over{\left(k^2+k_0^2\right)^{(d+\zeta)/2}}} \lb{Bespot} \ee
and $H(r)$ is given by the (for $d=2$, principal part) integral
\be H(r)= D\int d^d\bk\,\,{{1}\over{k^2}}\cdot {{e^{i\bk\bdot\br}}
           \over{\left(k^2+k_0^2\right)^{(d+\zeta)/2}}}, \lb{Wftn} \ee
so that $-\bigtriangleup H=K$. The scalar function $K(r)$ is essentially just
the standard Bessel
potential kernel \cite{ArS}, and may thus be expressed in terms of a modified
Bessel function:
\be K(r)=D {{2^{1-(\zeta/2)}k_0^{-\zeta}\pi^{d/2}}
           \over{\Gamma\left({{d+\zeta}\over{2}}\right)}}\cdot
                  (k_0r)^{\zeta/2}K_{\zeta/2}(k_0r). \lb{Besrep} \ee
The Hessian matrix $\partial_i\partial_jH(r)$ of the function $H$ of magnitude
$r=|\br|$ alone is
\be \partial_i\partial_jH(r)=\delta_{ij} J(r)+ \hr_i\hr_j\cdot
r{{dJ}\over{dr}}(r), \lb{Hess} \ee
with $J(r)= H'(r)/r$ and $\widehat{\br}=\br/r$. However, because ${\rm
Tr}\left(\grad\otimes\grad H\right)
= -K$, a Cauchy-Euler equation follows for $J(r)$:
\be r{{dJ}\over{dr}}(r)+ d\cdot J(r)= -K(r). \lb{CEeq} \ee
Due to the rapid decay of its Fourier transform, the function $J(r)$ is
continuous. Thus, the
relevant solution is found to be
\be J(r) = -r^{-d}\int_0^r \rho^{d-1} K(\rho) d\rho. \lb{AinV} \ee
in terms of $K(r)$. Using this expression for $J(r)$, along with
Eq.(\ref{Hess}), we thus find
\be D_{ij}(\br)= (K(r)+J(r))\delta_{ij}-(K(r)+d\cdot J(r))\hr_i\hr_j, \lb{Gexp}
\ee
which gives $D_{ij}$ as an explicit linear functional of $K$. Cf. \cite{MY},
equations (14.1),(14.3).

We are interested to consider the model in the range of length-scales $r\ll L$.
We require
some asymptotic expressions in that range:
\be K(\br) = K_{0} - K_{1} r^\zeta + O\left (k_0^2r^2\right ), \label{eq:A1}
\ee
with
\be K_0= D{{\Gamma\left({{\zeta}\over{2}}\right)\pi^{d/2}}
         \over{\Gamma\left({{d+\zeta}\over{2}}\right)}}\cdot k_0^{-\zeta}
\lb{Vzero} \ee
and
\be K_1=
D{{\Gamma\left({{\gamma}\over{2}}\right)\pi^{d/2}}\over{2^\zeta\cdot\zeta\cdot
                       \Gamma\left({{d+\zeta}\over{2}}\right)}}. \lb{Done} \ee
Also,
\be D_{ij}(\br) = D_{0}\delta_{ij} - D_{1}\cdot r^\zeta\cdot\left[\delta_{ij}
                 +{{\zeta}\over{d-1}}\left(\delta_{ij}-
{{r_ir_j}\over{r^2}}\right)\right]+ O\left (k_0^2r^2\right ) \label{eq:A2} \ee
with
\be D_0 = {{d-1}\over{d}} K_0 \lb{one} \ee
and
\be D_1 = {{d-1}\over{d+\zeta}} K_1 \lb{two} \ee
The first equation is derived by means of the known Frobenius series expansion
for the modified Bessel
functions (e.g. \cite{Erd}, section 7.2.2, equations (12),(13)). These give
\be z^\nu K_\nu(z)=
{{\Gamma(\nu)}\over{2^{1-\nu}}}-{{\Gamma(1-\nu)}\over{\nu\cdot
2^{1+\nu}}}z^{2\nu}
                                   +O\left(z^2\right). \lb{Frobser} \ee
{}From this expansion for $K_\nu(z)$ and from the representation (\ref{Besrep})
for $K(r)$ we obtain
the asymptotic expression (\ref{eq:A1}). Next, using (\ref{Gexp}), we observe
that, if $K(r)$ has
a power-law form, $K(r)= K r^\xi$, then it is easy to calculate that
\be D_{ij}(\br)= K r^\xi{{d-1}\over{d+\xi}}\left[\delta_{ij}+{{\xi}\over{d-1}}
                 \left(\delta_{ij}-\hr_i\hr_j\right)\right]. \lb{powfrm} \ee
Since $D_{ij}(r)$ is linearly related to $K(r)$, we may apply this formula to
the first
two terms in the expansion (\ref{eq:A1}), taking first $\xi=0$ and then
$\xi=\zeta$. This
yields the second expansion (\ref{eq:A2}).

As already shown by Kraichnan \cite{Kr68}, the equation for the $2$-point
correlation functions
in the model (\ref{2.1}) is closed. This was subsequently generalized to the
$N$-point
correlations \cite{Maj93,SS}. By now, these equations have been derived by
several arguments
and in many places, e.g. \cite{GK-I}. Therefore, we shall give no derivation
here. However,
 we make a few comments on the physical interpretation in the case $N=2$. Since
$D_{ij}(\bzed)
=D_0\delta_{ij}$, the two terms from $D_{ij}(\bzed)-D_{ij}(\br)$ may be treated
separately,
with the result that (\ref{1.3a}) may be written as
\be \partial_t\Theta(\br,t) = (D_0+2\kappa)\bigtriangleup\Theta(\br,t)
               -D_{ij}(\br)\grad_{\br_i}\grad_{\br_j}\Theta(\br,t). \lb{2.20}
\ee
Hence, we see that the first term gives essentially an augmentation of the
molecular
diffusivity, i.e. it produces an ``eddy diffusivity''
$\kappa_{eddy}={{1}\over{2}}D_0$.
In fact, it is the same eddy diffusivity which appears in the $N=1$ equation
\cite{GK-I}.
The second term, as we discuss in more detail below, represents additional
triadic interactions
between one velocity mode and two scalar modes. From (\ref{Vzero})-(\ref{two})
we see that both
of these terms are separately infrared divergent, in the limit $k_0\rightarrow
0$, but that
this infrared divergence cancels in the equation (\ref{1.3a}) for
$\Theta(\br,t)$. Since
we are only interested in the inertial-convective range behavior of the scalar,
when
$L_\theta(t)\ll L$, it is convenient for us to take the limit $k_0\rightarrow
0$.
This has been done in deriving (\ref{1.3}), which follows easily from
(\ref{1.3a}) and
(\ref{eq:A2}).

We shall also need below the equation for the spectral scalar energy transfer.
We introduce
the Fourier transform
\be \widehat{\Theta}(\bk,t):= {{1}\over{(2\pi)^d}}\int
\,\,\Theta(\br,t)e^{-i\bk\bdot\br}\,\,d\br.
    \lb{2.21} \ee
In terms of this, the scalar energy spectrum is
\be E_{\theta}(k,t):= {{1}\over{2}}\omega_{d-1} k^{d-1}\widehat{\Theta}(k,t),
\lb{2.22} \ee
where $\omega_{d-1}=2\pi^{d/2}/\Gamma\left({{d}\over{2}}\right)$ is the
$(d-1)$-dimensional
measure of the unit sphere in $d$-dimensions and, if the scalar statistics are
not isotropic,
$\widehat{\Theta}(k,t)$ is a spherical average. It is straightforward to
Fourier transform
(\ref{1.3a}), with the result
\begin{eqnarray}
\partial_t\widehat{\Theta}(\bk,t) & = & -k_ik_j \int d^d\bq
\,\,\widehat{D}_{ij}(\bq)
\left[\widehat{\Theta}(\bk,t)-\widehat{\Theta}(\bk-\bq,t)\right]
                   -2\kappa k^2 \widehat{\Theta}(\bk,t) \cr
                             \, & = & -(D_0+2\kappa) k^2
\widehat{\Theta}(\bk,t)+
        \int d^d\bq \,\,(\bk^\top\widehat{\bD}(\bq)\bk)\cdot
\widehat{\Theta}(\bk-\bq,t), \lb{2.23}
\end{eqnarray}
where $\widehat{D}_{ij}(\bq)= \widehat{K}(q)P_{ij}(\hat{\bq})$. Notice that
this last $d\times d$
matrix is positive semidefinite. Also, for each $\bk$,
$\widehat{\Theta}(\bk,t)\geq 0$, as a
statistical realizability requirement. Thus, we can now see that in the
spectral representation
for each wavevector $\bk$, the first $D$-term is always negative and represents
a ``loss'' term,
while the second $D$-term is always positive and gives a ``gain'' term. The
first is an ``eddy
diffusivity'' effect, as we have already discussed. The second can be seen to
result from triadic
interactions of a velocity mode with wavevector $\bq$ and two scalar modes with
wavevectors
$\bk-\bq$ and $\bk$. It is easy to derive from this an expression for the
transfer term $T_\theta(k,t)$
in the spectral energy balance equation
\be \partial_t E_\theta(k,t)= T_\theta(k,t)- 2\kappa k^2 E_\theta(k,t).
\lb{2.24} \ee
There are  ``loss'' and ``gain'' terms which are both infrared divergent in the
limits
$k_0\rightarrow 0$, but the infrared divergence cancels exactly in the equation
for $E_\theta(k,t)$.

\noindent {\it (2.2) Phenomenology of Turbulent Scalar Decay}

\noindent As discussed briefly in the Introduction, the standard phenomenology
of turbulent
decay is built upon two fundamental hypotheses: the {\it permanence of large
eddies} (PLE)
and {\it dynamic self-similarity} (DSS). We shall review each of these topics
here in turn,
commenting upon the original motivations for these hypotheses and their
dynamical justification
(or not) within the Kraichnan model.

First, we consider the permanence of large eddies. The motivation for this
hypothesis lies
in the phenomenon of {\it spectral backtransfer of energy}. As was first shown
by Proudman
and Reid in a calculation with the quasinormal closure for decaying
three-dimensional,
homogeneous turbulence \cite{PR}, the rate of change of the energy spectrum
asymptotically
at very low wavenumbers is dominated by a small but significant source of
energy, which arises
from nonlinear interactions of energy-range modes. Because the source-term is
positive,
and hence opposite in sign to the forward-cascading transfer through the
inertial subrange,
this phenomenon is called ``backtransfer''. According to calculations within
spectral closures---
such as quasinormal closure or its more sophisticated descendants, such as
eddy-damped quasinormal
markovian (EDQNM) closure---the transfer rate $T(k,t)$ is a power-law form
$\dot{B}(t)k^{d+1}$ in $d$
space-dimensions. See \cite{PR} and also \cite{MY}, Sections 15.5-15.6. The
same phenomenon
for scalar transfer that $T_\theta(k,t)\sim \dot{B}_\theta(t)k^{d+1}$ at low
$k$ was subsequently found
by Reid using again the quasinormal closure \cite{Reid55}. See also \cite{LMC}
and \cite{MY}, Section 19.4.
These closure calculations lead one to expect that, if the initial scalar
spectrum has a power-law
form, $E_\theta(k,t_0)\sim A k^n$ for $kL_\theta(t_0)\ll 1$, then this state of
affairs will be preserved
in time for $n<d+1.$ Indeed, with the latter assumption, the initial spectrum
will dominate
the time-integral $\int_{t_0}^t ds\,\,T_\theta(k,s) \sim B_\theta(t)k^{d+1}$
asymptotically
for small enough $k$. Hence, one may expect that $E_\theta(k,t)\sim A_\theta
k^n$ for
$kL_\theta(t)\ll 1$ with a constant independent of time. This is the usual
statement of the
hypothesis of permanence of large eddies, in a spectral formulation.

One may also formulate a permanence hypothesis in physical space. Thus, in the
scalar case, one
may suppose that, if $\Theta(r,t)\sim A'_\theta r^{-p}$ for $r\gg L(t)$ at the
initial time $t=t_0$,
then this relation will persist with the same constant $A'_\theta$ at later
times $t>t_0$, at least
when $p<d+2$. In the case of velocity correlations for Navier-Stokes
turbulence, Proudman and Reid
showed that pressure forces induce a long-range power-law $r^{-(d+2)}$ at any
positive time, even if
such correlations are not present initially. This is the exact physical-space
analogue of the $k^{d+1}$
spectral backtransfer. Thus, when $p<d+2$ the correlations present initially
shall dominate for $r$
sufficiently large. Similar arguments apply to the scalar correlations.
Although the long-range
pressure forces drop out of the expression for the first time-derivative of the
scalar correlation,
they appear in the second- and higher-order derivatives. Thus, a similar
power-law decay is expected
there.

Although there is a formal correspondence between the spectral space and
physical space formulations
of the hypothesis of permanent large eddies, the two versions are {\it not}
equivalent, as has
been emphasized in \cite{GSAFT}. Formally (e.g. see \cite{Wong}, Ch.IX.6,
Theorem 4),
\be E_\theta(k,t)\sim A_\theta k^n\,\,\,\,{\rm for}\,\,\,\,kL_\theta(t)\ll 1
\lb{2.25} \ee
corresponds to
\be \Theta(r,t)\sim A'_\theta r^{-p}\,\,\,\,{\rm for}\,\,\,\,r\gg L_\theta(t),
\lb{2.26} \ee
with $p=n+1$ and
\be A_\theta' = 2^p
{{\Gamma\left({{p}\over{2}}\right)\Gamma\left({{d}\over{2}}\right)}
                 \over{\Gamma\left({{d-p}\over{2}}\right)}}
                 A_\theta. \lb{2.27} \ee
Even formally, one can see that there is a problem when $p-d=2m,$ an even,
nonnegative integer,
since then $A_\theta'=0$ given by the above formula, when $A_\theta$ is finite.
In fact, this
case corresponds to $\widehat{\Theta}(k,t)\sim (k^2)^m $ for $kL_\theta(t)\ll
1$, which is thus
analytic at small $k$. Hence, one may expect the decay of $\Theta(r,t)$ in
physical space to
be faster than any power, consistent with the vanishing of $A_\theta'$
in (\ref{2.27}). It is therefore
quite possible that $E_\theta(k,t)$ exhibits a power-law (with or without
permanent coefficient)
and that $\Theta(r,t)$ has no power-law behavior whatsoever.

A case in point is when the function $\Theta(r,t)$ is integrable, with a
nonzero integral,
\be 0< K(t):= {{1}\over{(2\pi)^d}}\int d\br\,\,\Theta(\br,t)< \infty. \lb{2.28}
\ee
This is consistent with rapid decay in physical space faster than any power,
e.g. exponential.
However, the energy spectrum in this case exhibits a power-law at low
wavenumber. In fact,
\be E_\theta(k,t)\sim A(t) k^{d-1} \lb{2.29} \ee
with
\be A(t):= {{1}\over{2}}\omega_{d-1}\cdot K(t), \lb{2.30} \ee
just using the definition of the energy spectrum. This is sometimes called an
``equipartition
spectrum'', because it represents an average modal energy which is the same for
each Fourier
mode $\bk$. Not only is there a power-law spectrum, but, in fact, the
coefficient $A(t)$
is independent of time $t$, i.e. spectral PLE holds. This is true because
$K(t)$ is known
to be a constant of the motion, called the {\it Corrsin invariant}. It was
first derived
for the true passive scalar by S. Corrsin in 1951 \cite{Corr1,Corr2}, who
employed the equation
\be \partial_t\Theta(r,t) = {{2}\over{r^{d-1}}}{{\partial}\over{\partial
r}}\left[r^{d-1}\left(
             B(r,t)+ \kappa {{\partial\Theta}\over{\partial
r}}(r,t)\right)\right], \lb{2.31} \ee
with
\be B(r,t):=
\widehat{\br}\bdot\langle\bv(\br,t)\theta(\br,t)\theta(\bzed,t)\rangle.
\lb{2.32} \ee
This equation plays the role of the von K\'{a}rm\'{a}n-Howarth equation for the
temperature
field and it is analogous to the equation (\ref{1.3}) in the Kraichnan model.
It is not hard
to see that this equation implies that $dK(t)/dt=0$ when
$B(r,t)=o\left(r^{-(d-1)}\right)$. See
\cite{MY}, Section 15.2. Furthermore, the quasinormal closure calculations
suggest that
the indirect effect of pressure forces in $\partial_t B(r,t)$ (i.e. in the
higher-order
time-derivatives) lead to a decay at least as
$B(r,t)=O\left(r^{-(d+1)}\right)$, provided
that the decay was no slower initially. See \cite{MY}, Section 15.6. The
induced power-law
decay in the scalar correlation itself is
$\Theta(r,t)=O\left(r^{-(d+2)}\right)$, which is
integrable and thus the Corrsin integral should still be finite.

All of these arguments, which are derived under some closure assumptions for
true passive
scalars, are derived more directly and convincingly in the Kraichnan model. For
example,
if initially there is a power-law decay of correlations in space as in
(\ref{2.26}) in
the Kraichnan model, then
\be \partial_t\Theta(r,t_0)\sim -D_1 p(d+\gamma-p)A'_\theta
r^{-(p+\gamma)}\,\,\,\,
                                            {\rm for}\,\,\,\,r\gg
L_\theta(t_0), \lb{2.33} \ee
since the operator on the RHS of (\ref{1.3}) is homogeneous degree $-\gamma$.
Since $\gamma>0$,
this decay is faster, and it does not seem possible for the initial power-law
to be upset
at large enough $r$. Notice that in the Kraichnan model---unlike for true
turbulent scalars---
there is no action of pressure forces on the scalar correlators whatsoever.
Hence, in
the Kraichnan model, spatial PLE should hold for {\it all} $p>0$ and not just
for $0<p<d+2$.
In the spectral formulation, however, there is such a restriction. It follows
from (\ref{2.23})
that in the limit as $k\rightarrow 0$,
\be \partial_t E_\theta(k,t)\sim  -(D_0+2\kappa)k^2 E_\theta(k,t)
                                           + \dot{B}_\theta(t)k^{d+1} \lb{2.34}
\ee
with
\be \dot{B}_\theta(t):= \int d^d\bq
\,\,(\widehat{\bk}^\top\widehat{\bD}(\bq)\widehat{\bk})\cdot
     \widehat{\Theta}(\bq,t)\geq 0 \lb{2.35} \ee
and $\widehat{\bk}$ any unit vector. Thus, the ``gain'' term provides exactly
the $k^{d+1}$
power-law backtransfer term expected at low wavenumber $k$. Hence, spectral PLE
should hold
when there is a power-law spectrum of the form (\ref{2.25}) initially, with
$n<d+1$. Finally,
the theorem on the Corrsin invariant is easily seen to follow by multiplying
(\ref{1.3})
by $r^{d-1}$ and integrating over $r$. The result follows as long as
$\Theta(r,t)$ is
$o\left(r^{-(d-\gamma)}\right)$ near $r=0$ and $r=\infty$. The moral is that
PLE, whether
in physical or spectral space, should hold when the necessary power-laws are
present initially.

The second major hypothesis invoked in the traditional theory of turbulence
decay is dynamic
self-similarity (DSS). Mathematically, for the scalar decay problem, it amounts
to the assumption
that $\Theta(r,t)$ can be reduced to a function of a single variable
$\Phi(\rho)$ through a suitable
choice of the length and scalar concentration scales $\ell=\ell(t)$ and
$\vartheta=\vartheta(t)$:
\be \Theta(r,t)= \vartheta^2(t)\Phi\left({{r}\over{\ell(t)}}\right). \lb{2.36}
\ee
The function $\Phi(\rho)$ is called the {\it scaling function}. There is an
equivalent
spectral space version in which one assumes that
\be E_\theta(k,t)= \vartheta^2(t)\ell(t)F(k\ell(t)) \lb{2.37} \ee
for a spectral scaling function $F(\kappa)$. The scale $\vartheta(t)$ can be
taken to be
the rms scalar intensity
$\vartheta(t):=[\langle\theta^2(t)\rangle]^{1/2}=[\Theta(\bzed,t)]^{1/2}.$
The length-scale can be taken to be, for example, the scalar integral length
$L_\theta(t)$.
Within the validity of the hypothesis, all other relevant lengths are either
$0$,$\infty$,
or differ merely by a constant factor. It is natural to take $\ell(t)\propto
L_\theta(t)$,
because we want to consider a limit in which all dissipative length-scales go
to zero and we
also want to capture the energetics of the decay.

Clearly, the justification of the DSS hypothesis is more difficult than that of
PLE. Mathematically,
it provides a natural simplifying assumption, but truth and simplicity need not
coincide.
Our main purpose here is to investigate the validity of DSS in the specific
context
of the Kraichnan model. Although the methods we use are very specific to the
model, they
allow us to draw some conclusions that reasonably apply to other problems. The
present work should
also provide a testbed for general frameworks of understanding DSS. For
example, it is possible
that dynamic renormalization group methods can provide a more universal
foundation \cite{Gold}.

\vspace{.3in}

\noindent {\it Note}: Hereafter energy spectra, integral lengths, etc. will
refer to the scalar field
only and not to the velocity field. Hence we drop the subscript $\theta$
without any possibility
of confusion.

\newpage

\section{Self-Similar Decay and Its Breakdown}

\noindent {\it (3.1) Derivation \& Solution of the Scaling Equation}

\noindent In this section we shall completely identify and classify all of the
self-similar
decay solutions in the Kraichnan model. Following the discussion in section
2.2, we look for
solutions to (\ref{1.3}) in the form
\be \Theta(r,t) = \vartheta^2(t) \Phi\left({{r}\over{L(t)}}\right). \lb{3.1}
\ee
As there, we choose
\be \vartheta^2(t)=\Theta(0,t) \lb{3.2} \ee
and thus
\be \Phi(0)=1 \lb{3.3} \ee
by definition. Subsituting the {\it Ansatz} (\ref{3.1}) into (\ref{1.3}) gives
\be \left({{2\dot{\vartheta}(t)}\over{D_1
\vartheta}(t)L^{\zeta-2}(t)}\right)\Phi(\rho) +
    \left({{-\dot{L}(t)}\over{D_1
L^{\zeta-1}(t)}}\right)\rho{{\partial\Phi}\over{\partial\rho}}(\rho)
     = {{1}\over{\rho^{d-1}}}{{\partial}\over{\partial\rho}}
     \left (\rho^{d+\zeta-1}{{\partial\Phi}\over{\partial\rho}}\right).
\lb{3.4} \ee
The only way that this can hold with $\Phi(\rho)$ a function independent of
time $t$ is if
\be  {{2\dot{\vartheta}(t)}\over{D_1 \vartheta}(t)L^{\zeta-2}(t)}= -\alpha
\lb{3.5} \ee
and
\be {{\dot{L}(t)}\over{D_1 L^{\zeta-1}(t)}}= \beta \lb{3.6} \ee
for some constants $\alpha,\beta$. Indeed, the Wronskian of the two functions
$\Phi(\rho)$ and $\rho\Phi'(\rho)$
in terms of the logarithmic variable $\xi=\ln\rho$ is
$\Phi^2(\xi){{d^2}\over{d\xi^2}}\ln\Phi(\xi)$,
so that they are linearly independent on any interval where $\Phi(\rho)\neq 0$
and is not a pure power-law.
In order that ${{dE}\over{dt}}(t)<0$, we must have $\alpha>0$ (which explains
our choice
of sign in (\ref{3.5})). We see that the second equation (\ref{3.6}) is
identical with (\ref{1.6})
postulated for the scalar integral length, up to the factor of $\beta$. The
solution is $L(t)=
\left[L_0^\gamma+\beta\gamma D_1(t-t_0)\right]^{1/\gamma}$. We may always
choose $\beta$ to be unity
by a suitable choice of lengthscale in the {\it Ansatz}. In other words, there
are many lengths growing
according to the equation (\ref{3.6}) with some $\beta$ (e.g. the integral
length), but we choose the one obeying
the equation with $\beta=1$. This amounts to a rescaling
$\beta\rightarrow\beta'=1, L\rightarrow L'=\beta^{-1/\gamma}L$.
This transformation takes also $\alpha\rightarrow \alpha'=\alpha/\beta.$ We
shall always assume hereafter that
this rescaling has been done and simply write $\beta'=1,L'=L,\alpha'=\alpha$.
Then we arrive at the equation
\be -\alpha\Phi - \rho{{\partial\Phi}\over{\partial\rho}}  =
{{1}\over{\rho^{d-1}}}{{\partial}\over{\partial\rho}}
     \left (\rho^{d+\zeta-1}{{\partial\Phi}\over{\partial\rho}}\right).
\lb{3.7} \ee
The first coefficient equation (\ref{3.5}) may then be written using the second
(\ref{3.6}) as
\be  {{2\dot{\vartheta}(t)}\over{\vartheta}(t)}=
-\alpha{{\dot{L}(t)}\over{L}(t)}, \lb{3.8} \ee
whose solution is
\be \vartheta^2(t) = A \cdot L^{-\alpha}(t) \lb{3.9} \ee
for some constant $A$ with dimensions $[\Theta\cdot L^\alpha]$. It is clear
that any value of the constant $A$ is possible,
because the equation (\ref{1.3}) is linear and homogeneous. However, for any
solution we are always free to chose units
of the scalar field (e.g. temperature scale) so that $A\equiv 1$. Then
(\ref{3.9}) is identical with (\ref{1.7}) obtained
from the PLE hypothesis, if we take $\alpha=p$. In fact, we can verify this
directly from equation (\ref{3.7}),
if we substitute the asymptotic formula $\Phi(\rho) \sim C \rho^{-p}$ for
$\rho\gg 1$. We obtain
\be (\alpha-p)\rho^{-p} \sim p(d+\gamma-p)\rho^{-(p+\gamma)} \lb{3.10} \ee
for large $\rho$. Since $\gamma>0$, the righthand side is asymptotically
negligible compared with the left,
and we see that $\alpha=p$. Of course, we do not mean to imply that PLE is
necessarily true for self-similar
solutions, either spatially or spectrally. What is shown above is only that,
{\it if} spatial PLE holds, then
we have the identification $\alpha=p$. We may now write out the equation for
the scaling function in final form, as
\be \rho^{\zeta}\Phi''(\rho)+[(d+\zeta-1)\rho^{\zeta-1}+\rho]\Phi'(\rho)+
\alpha\Phi(\rho)=0. \lb{3.11} \ee

This equation can be solved in terms of confluent hypergeometric functions by
means of the substitution
$x= -{{\rho^\gamma}\over{\gamma}}$. In fact,
making this substitution into (\ref{3.11}) gives:
\be x{{\partial^2\Phi}\over{\partial x^2}} + \left[
{{d}\over{\gamma}}-x\right]{{\partial\Phi}\over{\partial x}}
                   -{{\alpha}\over{\gamma}}\Phi =0. \lb{3.14} \ee
This is a second-order differential equation with coefficients linear in the
variable $x$. Any such equation
can be solved in terms of confluent hypergeometric functions. In fact,
(\ref{3.14}) is {\it Kummer's equation}
with $a=\alpha/\gamma,\,c=d/\gamma$ (\cite{Erd}, Ch.6). This equation has two
solutions which are traditionally
denoted $\Phi(a,c;x)$ and $\Psi(a,c;x)$. The former is an entire function of
$x$ defined by the power series
\be \Phi(a,c;x)= 1 + {{a}\over{c}} {{x}\over{1!}}+{{a(a+1)}\over{c(c+1)}}
{{x^2}\over{2!}}+\cdots \lb{3.15} \ee
The second can be defined as
\be \Psi(a,c;x) = {{\Gamma(1-c)}\over{\Gamma(a-c+1)}}\Phi(a,c;x) +
                   {{\Gamma(c-1)}\over{\Gamma(a)}}x^{1-c}\Phi(a-c+1,2-c;x)
\lb{3.16} \ee
when $c$ is not an integer, and otherwise by a limit of this expression
(\cite{Erd},section 6.5). We see that
for $c>1$ and $c$ non-integer, the second solution has a power-law divergence
at $x=0$, while for integer $c$ the
divergence is logarithmic (\cite{Erd},section 6.7.1). The Wronskian of the two
solutions is $-{{\Gamma(c)}\over{\Gamma(a)}}$,
(\cite{Erd}, section 6.7) which are thus independent for $a,c>0$. These
conditions are satisfied in our problem
since $c={{d}\over{\gamma}}>1$ and $a={{\alpha}\over{\gamma}}>0$. Thus, we may
summarize the main conclusion
of this section as follows: {\it The unique solution of the scaling equation
(\ref{3.11}) satisfying
the boundary condition $\Phi(0)=1$ is
\be \Phi(\rho)
=\Phi\left({{\alpha}\over{\gamma}},{{d}\over{\gamma}};
-{{\rho^\gamma}\over{\gamma}}\right). \lb{3.17} \ee
Then, $\Theta(r,t):=\vartheta^2(t)\Phi(r/L(t))$ with $\vartheta(t),L(t)$
solutions of (\ref{3.5}) and
(\ref{3.6}) for $\beta=1$} \footnote{If we had used the original
length-scale $L(t)$ satisfying (\ref{3.6}) in defining the scaling solution,
rather that $L'(t)$ with
$\beta$ set equal to 1, then the result would have been instead
$\Phi\left({{\alpha}\over{\beta\gamma}},
{{d}\over{\gamma}};-{{\beta}\over{\gamma}}\rho^\gamma\right).$} {\it is an
exact solution of (\ref{1.3}).}
Hence, we have identified all possible scaling solutions in the Kraichnan
model. As an interesting
historical note, we observe that von K\'{a}rm\'{a}n and Howarth already reduced
their scaling equation,
with neglect of triple correlations, to Whittaker's form of the confluent
hypergeometric equation
\cite{vKH}. Kraichnan found the {\it energy spectrum} of the white-noise model
to obey Kummer's equation
in the Batchelor limit $\zeta\rightarrow 2,\gamma\rightarrow 0$ \cite{Kr74}.
Also, in the steady-state
with random forcing and molecular diffusion, the solution of the 2-point
correlation was given
as a Kummer function by Chertkov et al. in \cite{CFKL}.

\noindent {\it (3.2) Asymptotic Behaviors \& Permanence of Large Eddies}

\noindent The behavior of $\Phi(\rho)$ for small and large $\rho$ can be
obtained from the known
asymptotics of the Kummer function. For small $\rho$ it follows from
(\ref{3.15}) that
\be \Phi(\rho) = 1 -{{\alpha}\over{d\gamma}} \rho^\gamma +O(\rho^{2\gamma}).
\lb{3.18} \ee
This is the scaling that we expect for a dissipative solution. Because the
Kummer function is entire,
the only singularity of the scaling function is at $\rho=0$, due to the
fractional power $\gamma$.
Hence, we may apply Theorem 4 of Ch.IX.6 of \cite{Wong} to obtain, for
$0<\gamma<2$, that
\be \widehat{\Phi}(\kappa) \sim {{\alpha}\over{d\gamma}}\cdot
{{2^\gamma}\over{\pi^{{{1}\over{2}}(d+2)}}}
\Gamma\left({{\gamma+2}\over{2}}\right)
\Gamma\left({{d+\gamma}\over{2}}\right)\sin\left({{\pi\gamma}
\over{2}}\right)k^{-(d+\gamma)} \lb{3.20} \ee
for $\kappa\rightarrow\infty$. Hence, from (\ref{3.18}) the spectral scaling
function goes as
\be F(\kappa) \sim\alpha\cdot  2^\gamma
{{\Gamma\left({{\gamma}\over{2}}\right)\Gamma\left({{d+\gamma}\over{2}}\right)}
\over{4\pi\Gamma\left({{d+2}\over{2}}\right)}}
\sin\left({{\pi\gamma}\over{2}}\right)\kappa^{-(1+\gamma)}
    \lb{3.21} \ee
for $\kappa\gg 1$. Of course, this is the spectral law one would expect in the
inertial-convective range of the
white-noise model. In fact, going back to dimensionful quantities, we get
\be E(k,t) \sim {{1}\over{2}}\alpha C(\gamma,d)\vartheta^2(t)L^{-\gamma}(t)
\kappa^{-(1+\gamma)} \lb{3.22} \ee
with $C(\gamma,d):=  2^\gamma
{{\Gamma\left({{\gamma}\over{2}}\right)\Gamma\left({{d+\gamma}\over{2}}\right)}
\over{2\pi\Gamma\left({{d+2}\over{2}}\right)}}
\sin\left({{\pi\gamma}\over{2}}\right)$ for $kL(t)\gg 1$. However,
referring to the equation (\ref{3.5}) and using the definition of the scalar
dissipation $\chi(t)= -{{d}\over{dt}}
\left({{1}\over{2}}\vartheta^2(t)\right)$, one finds that
\be E(k,t) \sim C(\gamma,d){{\chi(t)}\over{D_1}} k^{-(1+\gamma)} \lb{3.23} \ee
for $kL(t)\gg 1$. This is identical to the result that holds in the forced
steady state, with the same (universal)
value of the constant $C(\gamma,d)$. In fact, the corresponding spatial result
is
\be \Theta(r,t) \sim \vartheta^2(t) - {{\chi(t)}\over{2\gamma d\cdot D_1}}
r^{\gamma} \lb{3.24} \ee
for $r\ll L(t)$, which coincides exactly with the result in equation (1.19b) of
\cite{CFKL}.

The large $\rho$ behavior of $\Phi(\rho)$ is obtained from the asymptotics of
the Kummer function for
large negative arguments (\cite{Erd}, section 6.13.1):
\be \Phi(a,c;x) = {{\Gamma(c)}\over{\Gamma(c-a)}}(-x)^{-a}\left[ 1 +
O(|x|^{-1})\right] \lb{3.25} \ee
as ${\rm Re}\,x\rightarrow -\infty$. Thus, we see that
\be \Phi(\rho) \sim
\gamma^{\alpha/\gamma}{{\Gamma\left({{d}\over{\gamma}}\right)}
\over{\Gamma\left(-{{\nu}\over{\gamma}}\right)}}
                \rho^{-\alpha}  \lb{3.26} \ee
for $\rho\gg 1$. We have set
\be \nu:= \alpha-d  \lb{3.27} \ee
so that $\alpha= d+\nu$. We see that $\Phi(\rho)$ has the power-law form
presumed in the PLE hypothesis, at least
when $\nu \neq \gamma \ell,\,\,\ell=0,1,2,...$. Otherwise, the coefficient of
the above asymptotic expression
vanishes and it no longer gives the leading behavior at large $\rho$. We shall
examine that case in detail below.
However, for all but such exceptional $\alpha$ values the spatial PLE in fact
holds. This can be seen by returning
to the dimensionful variables and using (\ref{3.9}), which gives
\be \Theta(r,t) \sim
\gamma^{\alpha/\gamma}{{\Gamma\left({{d}\over{\gamma}}\right)}
                \over{\Gamma\left(-{{\nu}\over{\gamma}}\right)}} r^{-\alpha}
\lb{3.28} \ee
for $r\gg L(t)$. This verifies the spatial PLE, for the non-exceptional values
$\alpha \neq d +\gamma \ell,\,\,\ell=0,1,2,...$,
since the coefficient of the asymptotic power is explicitly independent of the
time.

Let us now consider the exceptional cases, $\alpha = d +\gamma
\ell,\,\,\ell=0,1,2,...$. It turns out that these
are given in terms of elementary functions. In fact, for $\ell=0,1,2,...$
\begin{eqnarray}
\Phi(c+\ell,c;-x) & = & {{1}\over{(c)_\ell x^{c-1}}}
{{d^\ell}\over{dx^\ell}}\left[x^{c+\ell-1} e^{-x}\right] \cr
            \, & = & {{\ell!}\over{(c)_\ell}} L^{c-1}_\ell(x) e^{-x} \lb{3.29}
\end{eqnarray}
where $(c)_\ell= c(c+1)\cdots (c+\ell-1)$ and $L^{c-1}_\ell(x)$ is the {\it
generalized Laguerre polynomial} of degree $\ell$.
The first line follows from \cite{Erd}, 6.9.2(36) and the Kummer
transformation, \cite{Erd}, 6.3(7). The second line follows either from the
Rodriguez formula for the Laguerre polynomial or from \cite{Erd}, 6.4(11) with
$a=c$.
Using (\ref{3.29}) we see that
\be \Phi_\ell(\rho):=
\Phi\left({{d}\over{\gamma}}+\ell,{{d}\over{\gamma}};
       -{{\rho^\gamma}\over{\gamma}}\right)\
     = {{\ell!}\over{\left({{d}\over{\gamma}}\right)_\ell}}
L^{{{d-\gamma}\over{\gamma}}}_\ell\left({{\rho^\gamma}\over{\gamma}}\right)
             e^{-{{\rho^\gamma}\over{\gamma}}}. \lb{3.30} \ee
Explicitly, the first few functions are, for $\ell=\nu=0$,
\be  \Phi_0(\rho) = e^{-{{\rho^\gamma}\over{\gamma}}}, \lb{3.31} \ee
for $\ell=1,\nu=\gamma$
\be  \Phi_1(\rho) =
\left[1-{{\rho^\gamma}\over{d}}\right]e^{-{{\rho^\gamma}\over{\gamma}}},
\lb{3.32} \ee
and for $\ell=2,\nu=2\gamma$,
\be  \Phi_2(\rho) =
\left[1-{{2\rho^\gamma}\over{d}}+{{\rho^{2\gamma}}\over{d(d+\gamma)}}\right]
                                    e^{-{{\rho^\gamma}\over{\gamma}}}.
\lb{3.33} \ee
We see that all of the scaling functions in the exceptional cases $\alpha = d
+\gamma \ell,\,\,\ell=0,1,2,...$
do not behave as power-laws at large $\rho$, but, instead, have a
stretched-exponential decay. Thus, spatial
PLE cannot hold.

The case $\ell=\nu=0$ when $\alpha=d$ corresponds to a finite, nonvanishing
Corrsin invariant $K$. In fact, the dimensionless version
\be \widetilde{K}= {{\omega_{d-1}}\over{(2\pi)^d}}\int_0^\infty
e^{-{{\rho^\gamma}\over{\gamma}}}
                \rho^{d-1} d\rho \lb{3.34} \ee
can be calculated explicitly by substituting $t=\rho^\gamma/\gamma$ to yield a
Gamma integral:
\be \widetilde{K}= {{ \gamma^{{{d}\over{\gamma}}-1}}\over{2^{d-1}\pi^{d/2}}}
{{\Gamma\left({{d}\over{\gamma}}\right)}\over
       {\Gamma\left({{d}\over{2}}\right)}}. \lb{3.35} \ee
Thus, the spectral scaling function has the expected equipartition power-law at
low-wavenumbers:
\be F(\kappa) \sim {{1}\over{2}} \omega_{d-1} \widetilde{K} \kappa^{d-1}=
       {{\gamma^{{{d}\over{\gamma}}-1}}\over{2^{d-1}}}
{{\Gamma\left({{d}\over{\gamma}}\right)}\over
       {\left[\Gamma\left({{d}\over{2}}\right)\right]^2}} \cdot\kappa^{d-1}.
\lb{3.36} \ee
Although spatial PLE does not hold, the spectral PLE is valid. Indeed, the
dimensionful Corrsin integral is
\be K(t)= \widetilde{K} \vartheta^2(t)L^d(t) = \widetilde{K} \lb{3.37} \ee
using (\ref{3.9}) for $\alpha=d$ (with our convention that $A\equiv 1$). The
integral is explicitly independent of time,
in agreement with the general theorem on invariance. As an aside, we remark
that the stretched
exponential solution was noted by Majda and Kramer \cite{MajKram}.
In fact, it was found for $\gamma=2/3$ already by Batchelor in 1952 in a
slightly different context
\cite{Batch}. He considered (\ref{1.3}) as an equation for the particle
pair-separation distribution, in which case
it is just the scale-dependent diffusion equation proposed by Richardson
\cite{Rich}. However, along with the
new physical interpretation, there are also different mathematical requirements
on the solutions than those we impose.
To represent a probability distribution function only positive solutions with
unit integral are admissable.

Let us investigate the validity of the spectral PLE more generally. It is clear
from (\ref{3.26}) that
the spectral scaling function $F(\kappa)$ is not $C^\infty$ at $\kappa=0$. In
fact, the Fourier transform
of the radially symmetric function $\Phi(\rho)$ is given by
\be
\widehat{\Phi}(\kappa)={{1}\over{(2\pi)^{d/2}\kappa^{(d-2)/2}}}\int_0^\infty
                 \rho^{d/2} \Theta(\rho)J_{(d-2)/2}(\kappa\rho)\,\,d\rho
\lb{3.37a} \ee
where $J_\nu(z)$ is the Bessel function \cite{StW}. Then, by using the
Frobenius series expansion
of the Bessel function (\cite{Erd},7.2.1(2)), one obtains the formal Taylor
series of $\widehat{\Phi}(\kappa)$ in powers of $\kappa^2$:
\be \widehat{\Phi}(\kappa)\sim \sum_{j=0}^\infty B_j \kappa^{2j}, \lb{3.37b}
\ee
with
\be B_j:= {{(-1)^j}\over{\pi^{d/2}2^{2j+d-1}
j!\Gamma\left({{d+2j}\over{2}}\right)}}
          \int_0^\infty \rho^{2j}\Phi(\rho)\rho^{d-1}\,d\rho. \lb{3.37c} \ee
Because of the power-law decay with exponent $-\alpha$ in (\ref{3.26}), one
expects that
$B_j$ diverges for $2j\geq \nu\neq \gamma\ell,\,\ell=0,1,2,...$. This will be
verified below.
Hence, derivatives $\widehat{\Phi}^{(r)}(0)$ of order $r\geq \nu/2$ do not
exist. However, it
seems reasonable to assume that $\kappa=0$ is the only singularity of
$\widehat{\Phi}(\kappa)$,
which is $C^\infty$ elsewhere. In that case, Theorem 4 of Ch.IX.6 of
\cite{Wong} can be employed
to infer that the following asymptotic expansion holds for $\nu\neq 2j,\gamma
\ell,\,j,\ell=0,1,2,...$
and for $\kappa\ll 1$
\be \widehat{\Phi}(\kappa) \sim
\sum_{j=0}^{\left[\!\left[{{\nu}\over{2}}\right]\!\right]}
      B_j \kappa^{2j}+ B(\nu)\kappa^\nu  \lb{3.38} \ee
with
\be  B(\nu)={{\gamma^{\alpha/\gamma}}\over{\pi^{d/2}2^\alpha}}
{{\Gamma\left(-{{\nu}\over{2}}\right)
\Gamma\left({{d}\over{\gamma}}\right)}\over
{\Gamma\left(-{{\nu}\over{\gamma}}\right)
\Gamma\left({{\alpha}\over{2}}\right)}}. \lb{3.38a} \ee
In the case $\nu=2j,\nu\neq \gamma\ell,\,j,\ell=0,1,2,...$, the second term
after the sum
is changed to $B_*(\nu)\kappa^\nu\log \kappa$, modified by a logarithm. The
summation itself
represents the Taylor polynomial of degree
$\left[\!\!\left[{{\nu}\over{2}}\right]\!\!\right]$
of the contribution to $\widehat{\Phi}(\kappa)$ analytic at $\kappa=0$. Its
coefficients $B_j$
may be obtained from the formula (\ref{3.37c}) using with $\beta=2j$ the
integral
\be \int_0^\infty
\rho^{\beta}\Phi\left({{\alpha}\over{\gamma}},{{d}\over{\gamma}};
     -{{\rho^\gamma}\over{\gamma}}\right)\,\,\rho^{d-1}\,d\rho =
    \left\{\begin{array}{ll}
{{\Gamma\left({{d+\beta}\over{\gamma}}\right)
\Gamma\left({{d}\over{\gamma}}\right)
\Gamma\left({{\nu-\beta}\over{\gamma}}\right)}
\over{\Gamma\left({{\alpha}\over{\gamma}}\right)
\Gamma\left({{-\beta}\over{\gamma}}\right)}}
\gamma^{{{d+\beta}\over{\gamma}}-1}
& {\rm for}
           \,\,\,\,\beta<\nu \cr
         \pm\infty & {\rm for}\,\,\,\,\beta\geq\nu,
           \end{array} \right. \lb{3.38b} \ee
valid for the nonexceptional values $\nu\neq \gamma\ell,\,\,\ell=0,1,2,...$.
The sign $\pm$ in the second
case of (\ref{3.38b}) is given by ${\rm
sgn}\,\Gamma\left(-{{\nu}\over{\gamma}}\right)$,
the same as the power-law tail in (\ref{3.26}). For the exceptional values
$\nu=\gamma\ell,\,\,
\ell=0,1,2,...$ the first line in (\ref{3.38b}) is valid in both cases
$\beta<\nu$ and $\beta\geq\nu$,
with the convention that
$\Gamma\left(-{{\beta}\over{\gamma}}+\ell\right)/\Gamma\left(
-{{\beta}\over{\gamma}}\right)=\left(-{{\beta}\over{\gamma}}+\ell-1\right)\dots
\left(-{{\beta}\over{\gamma}}+1\right)\left(-{{\beta}\over{\gamma}}\right)=
\left(-{{\beta}\over{\gamma}}\right)_\ell$. The proof of this integral formula
proceeds by making the
change of variables $t=\rho^\gamma/\gamma$ and using the Laplace transform of
the Kummer function
\begin{eqnarray}
\int_0^\infty e^{-st}t^{b-1}\Phi(a,c;-t)\,dt & = &
{{\Gamma(b)\Gamma(c)\Gamma(a-b)}\over{\Gamma(a)\Gamma(c-b)}}
(1+s)^{c-a-b}F\left(c-a,1-a;b-a+1;-s\right) \cr
\, &  & \,\,\,\,\,\,\,\,\,
+{{\Gamma(c)\Gamma(b-a)}\over{\Gamma(c-a)}}
s^{a-b}F\left(a,1+a-c;a-b+1;-s\right)
    \lb{3.38c}
\end{eqnarray}
which is given in terms of the hypergeometric function $F$ and valid for ${\rm
Re}s>0$. This formula
follows from \cite{Erd},6.10(5) and 2.10(4). Notice that for the exceptional
cases $a=c+\ell,\,\ell=
0,1,2,...$ the second term in (\ref{3.38c}) vanishes. The limit $s\downarrow 0$
may be obtained
recalling that $F(0)=1$, with the result given in (\ref{3.38b}). We see that in
general for $\nu>0$,
there are nonzero $B_j$ terms from the analytic contribution which dominate the
Fourier transform
at small $\kappa$. However, it is interesting to note that the lowest
coefficient $B_0$, which
coincides with the dimensionless Corrsin integral $\widetilde{K}$, always
vanishes for $\nu>0$.
Thus, the Fourier transform $\widehat{\Phi}(\kappa)$ always vanishes at
$\kappa=0$ for $\nu>0$
and the second singular term proportional to $\kappa^\nu$ is formally the
leading-order contribution
for all $\nu<2$.

We may now return to the question of spectral PLE. The asymptotic
low-wavenumber expansion for
the spectral scaling function, corresponding to (\ref{3.38}), is
\be F(\kappa) \sim \sum_{j=0}^{\left[\!\left[{{\nu}\over{2}}\right]\!\right]}
A_j \kappa^{d+2j-1}+
              A(\nu)\kappa^{\alpha-1}  \lb{3.39} \ee
valid for $\kappa \ll 1$, with $A_j:= {{1}\over{2}}\omega_{d-1}B_j$ and
\be A(\nu):={{\gamma^{\alpha/\gamma}}\over{2^\alpha}}
{{\Gamma\left(-{{\nu}\over{2}}\right)}
\over{\Gamma\left(-{{\nu}\over{\gamma}}\right)}} \cdot
       {{\Gamma\left({{d}\over{\gamma}}\right)}\over
{\Gamma\left({{d}\over{2}}\right)\Gamma\left({{\alpha}\over{2}}\right)}}.
\lb{3.39a} \ee
If we return to dimensionful variables and use again (\ref{3.9}), we obtain
\be E(k,t) \sim \sum_{j=0}^{\left[\!\left[{{\nu}\over{2}}\right]\!\right]}
            A_j L^{-(\nu-2j)}(t)\kappa^{d+2j-1}+ A(\nu) k^{\alpha-1} \lb{3.40}
\ee
for $kL(t)\ll 1$. Spectral PLE appears to hold, in the sense that the
coefficient $A(\nu)$ of the
singular term is explicitly time-independent, for all but possibly the
exceptional values $\nu=
2j,\gamma \ell,\,\,j,\ell=0,1,2,...$. Of course, the singular term is the
leading one for $\nu<2$
and then spectral PLE appears to hold in the standard sense.

However, this is wrong for an important reason. While $A(\nu)>0$ for
$-d<\nu<\gamma$, it becomes negative
at rather higher values of $\nu$: $A(\nu,d)<0$ for
$\gamma<\nu<2\min\{1,\gamma\}$! Thus, realizability
of the scaling solution is violated in this case. This is a crucial issue which
we have neglected up until
now. In fact, only solutions with positive spectra over the whole wavenumber
range are physically
admissable. Thus, we have reached one of the important conclusions of this
work:
{\it No self-similar decay is possible in the Kraichnan model with
$\gamma<\nu<2\min\{1,\gamma\},$
since in that range the scaling solution has a negative spectrum at low
wavenumbers.}
We have proved this subject to a single assumption, that the origin is the only
singular point
for the Fourier transform of the scaling function. The result is verified by
explicit computations
for the special case $\gamma=1$ in an Appendix. In fact, we shall prove below
that self-similar
decay occurs for no exponent $\nu>\gamma$. The physics of this phenomenon will
be discussed
in section 3.4 below.

\newpage

\noindent {\it (3.3) Realizability of the Scaling Solutions}

\noindent We must now examine more closely the issue of realizability. As
already shown by Kraichnan
\cite{Kr68}, the 2-point correlation of the statistical problem (\ref{1.1})
must satisfy the closed partial
differential equation (\ref{1.3}). However, there can be solutions of the PDE
which do {\it not}
correspond to any solution of the statistical problem. The necessary and
sufficient condition
for a solution of (\ref{1.3}) to be realized as a solution of the statistical
problem (\ref{1.1})
is that it be positive-definite. Necessity is obvious. To see sufficiency, take
for any
positive-definite initial data $\Theta(r,0)$ the Gaussian measure $\mu_0$ over
scalar fields
$\theta_0(\br)$ with zero mean and with the given positive-definite function as
its 2-point covariance.
Then, the solution $\Theta(r,t)$ of the PDE (\ref{1.3}) with the specified
initial datum $\Theta(r,0)$
will be the same, by Kraichnan's result, as the 2-point correlation of the
statistical problem posed
by (\ref{1.1}) with random initial data $\theta_0(\br)$ distributed according
to $\mu_0$. It is a
corollary of this remark that the PDE (\ref{1.3}) is {\it positive-definiteness
preserving}, that is,
positive-definite solutions result from positive-definite initial data. The
upshot is that not
every scaling solution of (\ref{1.3}) that we found in Section 3.1 can
necessarily be realized
as a solution of the Kraichnan model (\ref{1.1}). This will only be true if it
is positive-definite,
and this question must now be addressed.

First, we give a general proof of realizability when $0<\alpha\leq d$
or $-d<\nu\leq 0$. We must show that $\Phi(\rho)$ is positive-definite as a
function on $d$-dimensional Euclidean
space. This is proved in two steps. We observe first that the $\nu=0$ function
$\Phi\left({{d}\over{\gamma}},{{d}\over{\gamma}};
-{{\rho^\gamma}\over{\gamma}}\right)= e^{-{{\rho^\gamma}\over{\gamma}}}$ is
positive-definite for $0<\gamma\leq 2$.
In fact, for $0<\gamma\leq 2$ the functions $e^{-{{\rho^\gamma}\over{\gamma}}}$
are the characteristic
functions (Fourier transforms) of positive probability densities, the
spherically symmetric stable
distributions of parameter $\gamma$. General multivariate stable laws were
first investigated by L\'{e}vy
\cite{Levy} and Feldheim \cite{Feld}; for an introduction to their basic
theory, see, for example,
the monograph of Zolotarev \cite{Zol}, Section I.6. The positive-definiteness
of the characteristic
functions $e^{-{{\rho^\gamma}\over{\gamma}}}$ for the spherically symmetric
stable distributions can
be obtained by an easy modification of the proof of Bochner for the
1-dimensional case \cite{Boch}.
However, with this result, we may then use a standard integral representation
for the Kummer functions
\be \Phi(a,c;x) = {{\Gamma(c)}\over{\Gamma(a)\Gamma(c-a)}} \int_0^1 e^{ux}
u^{a-1}(1-u)^{c-a-1} du \lb{3.41} \ee
valid for ${\rm Re}\,c>{\rm Re}\,a>0$. See \cite{Erd}, 6.5(1). From this we see
that
\be \Phi\left({{\alpha}\over{\gamma}},{{d}\over{\gamma}};
-{{\rho^\gamma}\over{\gamma}}\right)=
{{\Gamma\left({{d}\over{\gamma}}\right)}
\over{\Gamma\left({{\alpha}\over{\gamma}}\right)
     \Gamma\left(-{{\nu}\over{\gamma}}\right)}}
    \int_0^1 e^{-u{{\rho^\gamma}\over{\gamma}}}
    u^{{{\alpha}\over{\gamma}}-1}(1-u)^{-{{\nu}\over{\gamma}}-1} du, \lb{3.42}
\ee
when $-d<\nu<0$. Hence, $\Phi\left({{\alpha}\over{\gamma}},{{d}\over{\gamma}};
-{{\rho^\gamma}\over{\gamma}}\right)$
is a convex combination of positive-definite functions, in fact, averaged over
a Beta distribution. Thus, this scaling
function is always positive-definite for $-d<\nu\leq 0$.

We now show that the scaling functions are positive-definite---and thus
realizability holds---for the larger range
$-d<\nu\leq \gamma$ or $0<\alpha\leq d+\gamma$. As we have seen above, the
low-wavenumber spectrum remains positive
in the range $-d<\nu<\gamma$. Furthermore, we prove now that, if the scaling
function for $\nu=\gamma$ is positive
definite, then so are all the functions for $-d<\nu<\gamma$. This follows from
the identity
\footnote{Although it is not hard to prove, we did not find this formula in
standard treatises
on Kummer functions. We would be grateful for any reference.}
\be \Phi(c+\ell,c;x)={{\Gamma(c+1)}\over{\Gamma(c+\ell)
\Gamma(1-\ell)}}\int_0^1\Phi(c+1,c;ux)u^{c+\ell-1}(1-u)^{-\ell} du
\lb{3.43} \ee
for $1>{\rm Re}\,\ell> -{\rm Re}\,c$. This is most easily proved by expanding
both sides in a power-series in $x$ and
comparing the coefficients. One may also give a proof based upon \cite{Erd},
6.4(12) and 6.5(1). Thus,
\be \Phi\left({{d+\nu}\over{\gamma}},{{d}\over{\gamma}};
-{{\rho^\gamma}\over{\gamma}}\right)=
{{\Gamma\left({{d+\gamma}\over{\gamma}}\right)}
     \over{\Gamma\left({{d+\nu}\over{\gamma}}\right)
     \Gamma\left({{\gamma-\nu}\over{\gamma}}\right)}}
    \int_0^1 \Phi\left({{d+\gamma}\over{\gamma}},{{d}\over{\gamma}};
-u{{\rho^\gamma}\over{\gamma}}\right)
    u^{{{d+\nu}\over{\gamma}}-1}(1-u)^{-{{\nu}\over{\gamma}}} du, \lb{3.44} \ee
when $-d<\nu<\gamma$. Hence, if the scaling function  for $\nu=\gamma$,
$\Phi\left({{d+\gamma}\over{\gamma}},{{d}\over{\gamma}};
-{{\rho^\gamma}\over{\gamma}}\right),$ is positive-definite, then, as weighted
integrals of it with respect to a Beta distribution,
so are the scaling functions for $-d<\nu<\gamma$.

It therefore becomes important to answer whether
$\Phi_1(\rho)=\Phi\left({{d}\over{\gamma}}+1,{{d}\over{\gamma}};
-{{\rho^\gamma}\over{\gamma}}\right)$ is positive-definite or not. We shall
show that this question is related to
properties of the equipartition solution $\Phi_0(\rho)=
\Phi\left({{d}\over{\gamma}},{{d}\over{\gamma}};
-{{\rho^\gamma}\over{\gamma}}\right).$ In fact, it is not hard to show by
direct calculation that
\be d\cdot \Phi_1(\rho) = \left[d+\rho{{d}\over{d\rho}}\right]\Phi_0(\rho).
\lb{3.45} \ee
This is actually a particular example of a standard relation between Kummer
functions, \cite{Erd}, 6.4(11),
transformed by the substitution $x= -\rho^\gamma/\gamma$. What is useful here
is that the righthand side of
(\ref{3.45}) above is also just minus the lefthand side of (\ref{3.7}), in the
equipartition case $\alpha=d$.
However, the expression in (\ref{3.7}) arose by differentiating the scaling
{\it Ansatz} with respect to time
and dividing by $D_1 L^{-\gamma}(t)$. Thus, we infer that
\be \Phi_1(\rho) = {{-{{d}\over{dt}}\Theta_0(r,t)}\over{d\cdot D_1
L^{-\gamma}(t)}}, \lb{3.46} \ee
where $\Theta_0(r,t)=\vartheta^2(t)\Phi_0(r/L(t))$. This relation may be
Fourier-transformed easily, giving
\be \widehat{\Phi}_1(\kappa) =
{{-{{d}\over{dt}}\widehat{\Theta}_0(k,t)}\over{d\cdot D_1 L^{-\gamma}(t)}},
\lb{3.47} \ee
If we employ the relation $\widehat{\Theta}_0(k,t)=\widehat{\Phi}_0(kL(t))$,
which follows using (\ref{3.9}) for $\alpha=d$,
and (\ref{3.6}) for the time-derivative of $L(t)$, then (\ref{3.47}) yields the
final result
\be \widehat{\Phi}_1(\kappa) = -{{1}\over{d}}
\kappa{{d\widehat{\Phi}_0}\over{d\kappa}}(\kappa). \lb{3.48} \ee
{}From this we can see that realizability holds for the threshold case
$\ell=1,\nu=\gamma$ when the Fourier transform
$\widehat{\Phi}_0(\kappa)$ is a monotone nonincreasing function of spectral
radial coordinate $\kappa$.

To complete the proof, we must verify this property. In fact, the monotone
nonincreasing of the density
with respect to the radial cooordinate is equivalent---for spherically
symmetric functions---to the
property of {\it unimodality} of a multivariate density, as it has been defined
by Olshen and Savage
\cite{OlSa}. For a general introduction to the subject of unimodality and to
its proof for stable
probability laws in particular, see \cite{Zol}, Section 2.7, and \cite{DJ-D}.
The subject has a
rather colorful history, involving a series of published false proofs and
claims by eminent
mathematicians (including Kolmogorov), which is summarized in those works. The
first proof
of unimodality of the symmetric, one-dimensional stable distributions was given
by Wintner
in 1936 \cite{Wint}. It was then widely conjectured that {\it all}
one-dimensional stable
distributions are unimodal, but it took over forty years until a correct proof
was found
in 1978 by Yamazato \cite{Yama}. The proof of unimodality of spherically
symmetric stable distribution
functions in multi-dimensions was given about the same time, by S. J. Wolfe
\cite{Wolf}.
This is exactly the property we need to guarantee realizability for the
threshold case $\nu=\gamma$,
and thence, by equation (\ref{3.44}), for all $\nu$ in the range $-d<\nu\leq
\gamma$.

On the other hand, it is reasonable to conjecture that no scaling solutions
with $\nu>\gamma$ are realizable.
We have already established that realizability fails when $\gamma<\nu<
2\min\{1,\gamma\}$, by showing that the
low-wavenumber spectrum becomes negative. The same argument does not work for
all $\nu>\gamma$, because
(i) the coefficient of the singular contribution to the spectrum in
(\ref{3.39}) oscillates in sign as
$\nu$ is increased and (ii) the singular term is not the leading-order term at
low-wavenumbers for
$\nu>2$. However, we shall now prove that the scaling solutions for
$\nu>\gamma$ are indeed non-realizable.
The proof is based upon the fact that, for $\nu>\gamma$, two integrals vanish,
namely:
\be \int_{{\Bbb R}^d} \Phi(\rho) \,d^d\borho=\int_{{\Bbb R}^d}\rho^\gamma
\Phi(\rho) \,d^d\borho=0. \lb{add1} \ee
This is a direct consequence of the general formula (\ref{3.38b}) for the case
$\beta=\gamma$.
We shall now show (following a suggestion of D. Thomson \cite{Thom1,Thom2})
that these two conditions
are equivalent to
\be \int_{{\Bbb R}^d}\kappa^{-(d+\gamma)} \widehat{\Phi}(\kappa)
\,d^d\bokappa=0, \lb{add2} \ee
when realizability (or nonnegativity of $\widehat{\Phi}$ pointwise) is assumed.
To prove (\ref{add2})
we make use of the following generating functional for the moment-integrals in
(\ref{add1}):
\be G(t):= {{1}\over{(2\pi)^d}}\int_{{\Bbb R}^d}
e^{-t{{\rho^\gamma}\over{\gamma}}}\Phi(\rho) \,d^d\borho. \lb{add3} \ee
Then it is easy to see that (\ref{add1}) is equivalent to $G(0)=\dot{G}(0)=0$
or, as well,
$\lim_{t\rightarrow 0}{{G(t)}\over{t}}=0$. Next we make use of Parseval's
theorem to rewrite (\ref{add3}) as
\be G(t):= \int_{{\Bbb
R}^d}{{1}\over{t^{d/\gamma}}}\widehat{\Phi}_0
           \left({{\kappa}\over{t^{1/\gamma}}}\right)
           \widehat{\Phi}(\kappa) \,d^d\bokappa, \lb{add4} \ee
where $\widehat{\Phi}_0(\kappa)$ is the Fourier transform of the scaling
function for $\nu=0$,
i.e. the density of the multidimensional L\'{e}vy stable distribution with
parameter $\gamma$.
This step is justified because, clearly, $\Phi_0\in L^2$, and because, for
general $\nu>0$,
boundedness and the large-$\rho$ decay in (\ref{3.26}) imply that as well
$\Phi\in L^2$.
Next we observe from (\ref{3.20}) that $\widehat{\Phi}_0(\kappa)\sim
c\cdot\kappa^{-(d+\gamma)}$
as $\kappa\rightarrow\infty$, for some positive constant $c>0$. Thus,
\be \lim_{t\rightarrow
0}{{1}\over{t^{(d+\gamma)/\gamma}}}
\widehat{\Phi}_0\left({{\kappa}\over{t^{1/\gamma}}}\right)
     = c \cdot\kappa^{-(d+\gamma)}. \lb{add5} \ee
We see finally, by Fatou's lemma, that
\begin{eqnarray}
0 = \liminf_{t\rightarrow 0} {{G(t)}\over{t}} & = &
\liminf_{t\rightarrow 0} \int_{{\Bbb R}^d}{{1}\over{t^{(d+\gamma)/\gamma}}}
\widehat{\Phi}_0\left({{\kappa}\over{t^{1/\gamma}}}\right)
\widehat{\Phi}(\kappa)\,d^d\bokappa \cr
\, & \geq &
\int_{{\Bbb R}^d}\liminf_{t\rightarrow 0} {{1}\over{t^{(d+\gamma)/\gamma}}}
\widehat{\Phi}_0\left({{\kappa}\over{t^{1/\gamma}}}\right)
\widehat{\Phi}(\kappa)\,d^d\bokappa \cr
\, & =   & c\int_{{\Bbb R}^d}\kappa^{-(d+\gamma)} \widehat{\Phi}(\kappa)
\,d^d\bokappa \geq 0, \lb{add6}
\end{eqnarray}
where the last inequality holds by the assumed nonnegativity of
$\widehat{\Phi}(\kappa)$. Thus, the identity (\ref{add2})
is established. However, it follows immediately then that
$\widehat{\Phi}(\kappa)=0$ for $a.e.\,\,\kappa$, which is a clear
contradiction. Thus, the assumption that $\widehat{\Phi}(\kappa)\geq 0$ for all
$\kappa$ cannot be correct.

We may summarize the conclusions of this section as follows: {\it The scaling
functions
$\Phi(\rho)=\Phi\left({{\alpha}\over{\gamma}},{{d}\over{\gamma}};
-{{\rho^\gamma}\over{\gamma}}\right)$
are positive-definite, radially-symmetric functions on ${\Bbb R}^d$ for all
$\alpha$ in the
interval $(0,d+\gamma]$ and for no $\alpha$ in the range $(d+\gamma,\infty)$.
In particular, the scaling solutions
$\Theta(r,t)=\vartheta^2(t)\Phi(r/L(t))$ for $\alpha\leq d+\gamma$ are all
realized
as solutions of the statistical problem posed by the equation (\ref{1.1}) with
the random
initial data $\mu_0$.} It is interesting to note that, for cases $\nu>\gamma$
when the
low-wavenumber spectrum is positive, there must be at least one
negative spectral interval away from the origin. This is explicitly
verified in the Appendix for $\gamma=1$, in which case the spectral
scaling functions $F(\kappa)$ can be calculated in a closed form
(related to the multidimensional Cauchy distribution) for all of the
exceptional values
$\nu=\ell,\,\,\ell=0,1,2,...$.

\newpage

\noindent {\it (3.4) A Physical Explanation of the Results}

\noindent We have now shown there are realizable self-similar solutions for
$-d<\nu\leq \gamma$ but not
for $\nu>\gamma$. Although $\nu=\gamma$ corresponds to a realizable DSS
solution,  something a bit strange
must occur in that case. For example, PLE cannot hold either spatially or
spectrally. If there is a power-law in
$E_1(k,t)$ at low-wavenumber, it must be distinct from the power
$k^{d+\gamma-1}$ which is naively expected. Indeed,
if that naive power-law occurred, then it would imply a corresponding spatial
decay $r^{-(d+\gamma)}$, since $\gamma$
lies in the range $0<\gamma<2$ where this deduction is correct. However, we
know that $\Theta_1(r,t)$ decays faster than
{\it any} inverse power of $r$. Hence, if any power-law at all occurs in the
spectrum, it must be different from the naive
one and, in fact, it must correspond to $\nu$ an even, positive integer. We
show now that there is a spectral power-law
$k^{d+1}$ (naively corresponding to $\nu=2$) and representing scalar
backtransfer.

More precisely, we show that
\be \widehat{\Phi}_1(\kappa) \sim
{\gamma^{{{d+\zeta}\over{\gamma}}}\over{d(4\pi)^{d/2}}}
{{\Gamma\left({{d+2}\over{\gamma}}\right)}
\over{\Gamma\left({{d+2}\over{2}}\right)}} \kappa^2 \lb{3.49} \ee
for $\kappa\ll 1$. To prove this, we use (\ref{3.47}) again, but calculate now
the time-derivative using the
transfer equation discussed in Section 2.1:
\be \partial_t\widehat{\Theta}_0(k,t) = -k_ik_j \int d^d\bq
\,\,\widehat{D}_{ij}(\bq)
\left[\widehat{\Theta}_0(k,t)-\widehat{\Theta}_0(|\bk-\bq|,t)\right], \lb{3.50}
\ee
where $\widehat{D}_{ij}(\bq)= \widehat{K}(q)P_{ij}(\hat{\bq})$. We can take the
limit as $k\rightarrow 0$
in this expression, with the result that
\be \partial_t\widehat{\Theta}_0(k,t) \sim -{{d-1}\over{d}} k^2 \int d^d\bq
\,\,\widehat{K}(q)
\left[\widehat{\Theta}_0(0,t)-\widehat{\Theta}_0(q,t)\right], \lb{3.51} \ee
where the factor $(d-1)/d$ comes from the angular average of
$P_{ij}(\hat{\bq})$. Fourier inversion gives
\be \int d^d\bq
\,\,\widehat{K}(q)\left[\widehat{\Theta}_0(0,t)-\widehat{\Theta}_0(q,t)\right]
     = {{1}\over{(2\pi)^d}}\int d^d\br \,\,[K(0)-K(r)]\Theta_0(r,t). \lb{3.52}
\ee
In this form, the limit as $k_0\rightarrow 0$ is easy to evaluate, with the
result
\be \lim_{k_0\rightarrow 0}\int d^d\br \,\,[K(0)-K(r)]\Theta_0(r,t)= K_1 \int
d^d\br \,\,r^\zeta\Theta_0(r,t), \lb{3.53} \ee
for $K_1= \left({{d+\zeta}\over{d-1}}\right) D_1.$ By substituting the scaling
{\it Ansatz} and using
again (\ref{3.9}), one finds
\be \int d^d\br \,\,r^\zeta\Theta_0(r,t) =
                   L^{\zeta}(t)\cdot\omega_{d-1}\int_0^\infty
\,\,\rho^{d+\zeta-1}\Phi_0(\rho)\,d\rho. \lb{3.54} \ee
The right side of (\ref{3.54}) can be reduced to a Gamma integral by the
subsitution $t= \rho^\gamma/\gamma$, giving
\be \int_0^\infty \,\,\rho^{d+\zeta-1}\Phi_0(\rho) \,d\rho =
\gamma^{{{d+\zeta}\over{\gamma}}-1}
\Gamma\left({{d+\zeta}\over{\gamma}}\right). \lb{3.55} \ee
Putting together (\ref{3.51})-(\ref{3.55}) gives finally
\begin{eqnarray}
\partial_t\widehat{\Theta}_0(k,t) & \sim &
-{{D_1}\over{(2\pi)^d}}\left({{d-1}\over{d}}\right)
\left({{d+\zeta}\over{d-1}}\right)\omega_{d-1}
\gamma^{{{d+\zeta}\over{\gamma}}-1}
\Gamma\left({{d+\zeta}\over{\gamma}}\right) L^\zeta(t)(\kappa/L(t))^2 \cr
                               \, &   =  & - D_1 L^{-\gamma}(t)\cdot
\gamma^{{{d+\zeta}\over{\gamma}}}
         \Gamma\left({{d+2}\over{\gamma}}\right)\kappa^2
\left/(4\pi)^{d/2}\Gamma\left({{d+2}\over{2}}\right)\right. . \lb{3.56}
\end{eqnarray}
Substituting this into (\ref{3.47}) yields (\ref{3.49}), as claimed.

The result may be easily understood in terms of our general result (\ref{3.38})
for the low-wavenumber
asymptotics. The coefficient $B(\nu)$ of the singular contribution vanishes at
$\nu=\gamma$, because
of the Gamma function $\Gamma\left(-{{\nu}\over{\gamma}}\right)$ in the
denominator of (\ref{3.38a}).
Thus, the leading term ought to be $B_1 \kappa^2$. If we substitute
$\beta=2,\nu=\gamma$ in (\ref{3.38b}),
the integral is found to be
$-{{2}\over{d}}\Gamma\left({{d+2}\over{\gamma}}\right)
\gamma^{{{d+2}\over{\gamma}}-1}$. This yields
$B_1={\gamma^{{{d+\zeta}\over{\gamma}}}\over{d(4\pi)^{d/2}}}
{{\Gamma\left({{d+2}\over{\gamma}}\right)}
\over{\Gamma\left({{d+2}\over{2}}\right)}}$ in (\ref{3.37c}),
in exact agreement with (\ref{3.49}).

In terms of the spectral scaling function, the result for our DSS solution is
\be F_1(\kappa) \sim
{{\gamma^{{{d+2}\over{\gamma}}-1}}\over{2^{d+1}}}
{{\Gamma\left({{d+2}\over{\gamma}}\right)}
                \over{\left[\Gamma\left({{d+2}\over{2}}\right)\right]^2}}
\cdot\kappa^{d+1} \lb{3.57} \ee
for $\kappa\ll 1$. Although self-similar, spectral PLE must be violated in this
solution. In fact, when the naive power
is replaced by any other, the DSS solution will automatically develop a
time-dependent low-wavenumber coefficient, in contrast
to (\ref{3.40}). To see this, we may return to dimensionful variables in
(\ref{3.57}), using now (\ref{3.9}) with
$\alpha=d+\gamma$. The result is
\be E_1(k,t) \sim A(\gamma,d)L^\zeta(t) k^{d+1} \lb{3.58} \ee
for $k L(t)\ll 1$, with
$A(\gamma,d):={\gamma^{{{d+2}\over{\gamma}}-1}\over{2^{d+1}}}
{{\Gamma\left({{d+2}\over{\gamma}}\right)}
\over{\left[\Gamma\left({{d+2}\over{2}}\right)\right]^2}}$. Because of the
``leftover'' factor of $L^\zeta(t)$,
the coefficient of the asymptotic $k^{d+1}$ power-law is now explicitly
time-dependent. In fact, it exhibits power-law
growth $L^\zeta(t)\sim [\gamma D_1(t-t_0)]^{\zeta/\gamma}$ for long times
$t-t_0\gg L^\gamma(t_0)/D_1$. This increase is consistent
with the interpretation of the low-wavenumber spectrum as arising from scalar
backtransfer.

Since there are no realizable self-similar solutions for $\nu>\gamma$, one is
led to the following
question: how shall an initial scalar spectrum $E(k,t_0)\sim A k^{\alpha_0-1}$
at small $k$ with
$\alpha_0=d+\nu_0$ decay asymptotically at long times? According to the
traditional view,
when $\nu_0\geq 2$, then the decay shall be asymptotically self-similar at long
times
described by the spectrum $E_1(k,t)$ for the threshold case $\nu=\gamma$ above.
We see no reason
to doubt the validity of this view. On the other hand, when $-d<\nu_0<2$, the
traditional view states
that the decay will be described at long times by the self-similar solution
with $\nu=\nu_0$.
This is perfectly consistent when $-d<\nu_0<\gamma$ and, again, we see no
reason to doubt the
traditional picture of the decay. However, when $\gamma<\nu_0<2$, then there is
no realizable
self-similar solution with $\nu=\nu_0$! Hence, the traditional view must be
wrong for $\gamma<\nu_0<2$.

We cannot from our present analysis say what happens in the case
$\gamma<\nu_0<2$, because it lies outside
consideration of the self-similar solutions themselves. We shall treat this
dynamical problem in
Section 4. However, we shall now show that the scenario proposed in
\cite{GSAFT} for the Burgers decay
gives a completely consistent account of the known facts also in the Kraichnan
model. The picture that
is proposed is of a {\it two-scale decay}. In addition to the integral length
$L(t)$, there is another
length-scale $L_*(t)\gg L(t)$, which separates an inner solution $E_{in}(k,t)$
for $kL_*(t)\gg 1$ and
an outer solution $E_{out}(k,t)$ for $kL_*(t)\ll 1$. The inner solution is just
the self-similar decay
solution $E_1(k,t)$ for $\nu=\gamma$. The outer solution is the same as the
initial spectrum
$E_{out}(k,t)\sim A k^{\alpha_0-1}$ for $kL_*(t)\ll 1$. Hence, spectral PLE
holds, but only
in the outer region. Now by matching the inner and outer solutions, one can
find the crossover
length-scale $L_*(t)$, or, equivalently, its associated wavenumber $k_*(t)$, as
\be A [k_*(t)]^{d+\nu_0-1} \sim D(t-t_0)^{\zeta/\gamma} [k_*(t)]^{d+1},
\lb{3.59} \ee
or
\be L_*(t) \sim (t-t_0)^{{\zeta}\over{(2-\nu_0)\gamma}}. \lb{3.60} \ee
Since $L(t)\sim (t-t_0)^{1/\gamma}$, the inequality $L_*(t)\gg L(t)$ necessary
for validity
of this picture only holds if $\zeta/(2-\nu_0)>1$ or if $2>\nu_0>\gamma$. It is
a highly nontrivial
test of consistency that the critical value $\nu_0=\gamma$, above which the
ratio $R(t):=L_*(t)/L(t)$
grows, coincides exactly with the value $\nu=\gamma$, above which there is no
realizable self-similar
decay solution. This gives us some confidence in the correctness of the picture
proposed. If this
picture is correct, then at very long times DSS is restored and the decay is
described by the
inner solution spectrum $E_1(k,t)$, since $L_*(t)\rightarrow +\infty$ in units
of $L(t)$.
As $\nu_0\rightarrow 2-$, the growth rate of $L_*(t)$ becomes infinitely fast
and the outer
solution region disappears, in agreement with the traditional view.  Thus, PLE
and DSS
both hold for $\gamma<\nu_0<2$, but the decay is not what one would naively
expect for
DSS+ PLE, because the outer range where PLE holds is not part of the
inner-range DSS solution.

A similar picture may be developed in physical space, but, as noted in
\cite{GSAFT},
the separation of the length-scales is not as sharp. Now one would expect
$\Theta_{in}(r,t)
=\Theta_1(r,t)$ for $r\ll L_*(t)$ and $\Theta_{out}(r,t)\sim A' r^{-(d+\nu_0)}$
for $r\gg L_*(t)$.
Note that for $r\approx L(t)$,
\be \Theta_{in}(r,t)\sim [L(t)]^{-(d+\gamma)}\gg \Theta_{out}(r,t)\sim
[L(t)]^{-(d+\nu_0)}. \lb{3.new} \ee
The crossover occurs at a larger length-scale $L_*(t)$, which is found by the
matching condition
\be [L(t)]^{-(d+\gamma)}\exp[-(1/\gamma)(L_*(t)/L(t))^\gamma]\sim
[L_*(t)]^{-(d+\nu_0)}.
     \lb{3.61} \ee
It is easy to see that this implies a solution for the ratio
$R_*(t)=L_*(t)/L(t)$ of the form
\be R(t)\sim [\log(t-t_0)]^{1/\gamma}. \lb{3.62} \ee
Hence, $L_*(t)$ is only larger than $L(t)$ by a logarithmic term in physical
space.

\newpage

\section{Convergence to Self-similar Solutions}

\noindent {\it (4.1) The Long-time Scaling Limit}

\noindent In this section, we identify the time-dependent solutions of the
$\Theta$ equation
(\ref{1.3}) which show eventually a self-similar form of decay. That is, we
find
domains of attraction of the self-similar solutions constructed in Section 3.
We are interested in observing the solutions on a range of length-scales
comparable
to $L(t)$ and at a level of scalar amplitude comparable to the rms fluctuation
$\vartheta(t)$.
Hence, we consider the rescaled solutions
\be  \Theta(r,t) = \vartheta^{2}(t)\Gamma(r/L(t),\tau(t)), \label{C2} \ee
where ${\dot{\vartheta}\over \theta}= - {\alpha \over 2}{\dot{L}\over L}$,
${\dot{L}\over L}=D_1 L^{\zeta -2}$ are as in (\ref{3.5}),(\ref{3.6}) and
\be \tau(t): = \log L^\gamma \sim \log (t-t_0). \lb{C1} \ee
The function $\Gamma$ solves:
\be
 \gamma {{\partial
\Gamma}\over{\partial\tau}}(\rho,\tau)={{1}\over{\rho^{d-1}}}
{{\partial}\over{\partial\rho}}\left[ \rho^{d+\zeta -1} {{\partial
\Gamma}\over{\partial\rho}}
 (\rho,\tau)\right] + \rho{{\partial \Gamma}\over{\partial\rho}}(\rho,\tau)
 +\alpha \Gamma(\rho,\tau), \label{C3}
\ee
where $\rho = r/L(t)$. This change of variables may be made for any $\alpha>0$,
and,
if we wish to make this explicit we shall refer to the above function as
$\Gamma_{(\alpha)}(\rho,\tau)$.
Of course, it is only possible that a nontrivial scaling limit is obtained as
$\tau\rightarrow\infty$ for one exponent $\alpha$. If a nontrivial limit is
obtained
for $\alpha$, then, for any other exponent, say, $\alpha'$, it follows that
\be    \lim_{\tau\rightarrow\infty}\Gamma_{(\alpha')}(\rho,\tau)= \left\{
       \begin{array}{ll}
        \infty & {\rm if}\,\,\,\,\alpha'>\alpha \cr
        0 & {\rm if}\,\,\,\,\alpha'<\alpha
       \end{array}\right. \lb{C3a} \ee
Of course, this means that if the scaling limit
$\Gamma_{(\alpha')}(\rho,\tau)\rightarrow 0$
(resp. $\infty$) for $\alpha'$, then an exponent $\alpha$ larger (resp.
smaller)
than $\alpha'$ is required for a nontrivial limit (if this is possible at all).

To analyze the equation (\ref{C3}), we make the change of variables $x =
\gamma^{-1}\rho^{\gamma}$,
which is the same as in Section 3 up to a sign. We obtain
\be {{\partial \Gamma}\over{\partial\tau}}(x,\tau)
     =x {{\partial^2\Gamma}\over{\partial x^2}}(x,\tau)
       + (c+x){{\partial\Gamma}\over{\partial x}}(x,\tau)
       + a \Gamma(x,\tau), \label{C4} \ee
with $a=\alpha/\gamma,c=d/\gamma$. Let $\Phi_{(\alpha)}(x):=\Phi(a,c;-x)$ be
the self-similar solution
discussed in Section 3, for the same choice of $\alpha$ as employed in
$\Gamma_{(\alpha)}$.
Then $\Phi$ is a steady state solution of (\ref{C4}). We define
$\Alpha(x,\tau)$ via
$\Gamma(x,\tau) = \Phi(x)\Alpha(x,\tau).$ Clearly, to establish convergence
$\Gamma(x,\tau)
\rightarrow \Phi(x)$, it is enough to show that $\Alpha(x,\tau)\rightarrow 1$.
We find for $\Alpha$ that:
\be {{\partial \Alpha}\over{\partial\tau}}(x,\tau)
     = x {{\partial^2\Alpha}\over{\partial x^2}}(x,\tau)
       + \left(c + x + 2 x {{\Phi'(x)}\over{\Phi(x)}}\right)
       {{\partial\Alpha}\over{\partial x}}(x,\tau) \label{C5} \ee
The transformation to $A(x,t)$ has removed the explicit $\alpha$ dependence,
which is now
represented only through $\Phi_{(\alpha)}$.

We shall now obtain characterizations of the domains of convergence of the
self-similar
solutions for each $\alpha$. In this analysis, it is important to distinguish
two general
classes of initial data: those of ``rapid decay'' for which
\be \int_{{\Bbb R}^d} |\Theta(r,0)|^2 e^{{{r^\gamma}\over{\gamma}}} d\br
<\infty \lb{C5zed} \ee
and those of ``slow decay'' for which the above integral is infinite. Thus, the
initial
data showing rapid decay belong to $L^2$ with a stretched-exponential weight
and must decay
at least as fast as the weight. This turns out to be a useful formal criterion
for ``rapid-decay''.
Initial data with power-law decay at large $r$---which are of particular
interest in view
of the question of validity of spatial PLE---are classified as ``slow decay''
functions.
Convergence results will be established for both classes of initial data below.

\noindent {\it (4.2) Initial Data with Rapid Decay}

\noindent We analyze first the case of ``rapid decay''. If we make the scaling
with $\alpha=d$
and define correspondingly $\Gamma(x,\tau)=A(x,\tau)\Phi_0(x)$, then (\ref{C5})
becomes
\be  {{\partial \Alpha}\over{\partial\tau}}(x,\tau)
     = x {{\partial^2\Alpha}\over{\partial x^2}}(x,\tau)
       + (c -x){{\partial\Alpha}\over{\partial x}}(x,\tau)
     := {\cal L}_0A(x,t) \label{CC1} \ee
We used $\Phi_0(x)=e^{-x}$. The operator ${\cal L}_0$ has as its eigenfunctions
the generalized
Laguerre polynomials $L_\ell^{c-1}(x)$ with eigenvalues $-\ell$. Indeed, the
Laguerre polynomial
is characterized as the unique solution $y=L_\ell^{c-1}(x)$ of the second-order
equation
\be xy''+ (c-x)y' + \ell y=0 \lb{CC2} \ee
which is regular at the origin. See \cite{Erd}, 6.9.2(36) or \cite{Szego}, 5.1,
5.3. In terms
of $A(x,0)$, the rapid decay criterion (\ref{C5zed}) becomes
\be \int_0^\infty |A(x,0)|^2 e^{-x}x^{c-1}dx <\infty \lb{CC3} \ee
It is known that the generalized Laguerre polynomials
$\left\{L_\ell^{c-1}(x):\ell=0,1,2,...\right\}$
are a complete, orthogonal set on the interval $(0,\infty)$ with the weight
$w_c(x)= x^{c-1}e^{-x}$.
See \cite{Szego}, Theorem 5.7.1. Thus, it follows from the above remarks that
the solution $A(x,\tau)$
of (\ref{CC1}) with initial datum $A(x,0)$ also satisfies the condition
(\ref{CC3}) and, furthermore,
has the expansion
\be A(x,\tau) = \sum_{\ell=0}^\infty a_\ell e^{-\ell\tau} L_\ell^{c-1}(x)
\lb{CC4} \ee
which converges in the $L^2$-sense with weight $w_c(x)$. The expansion
coefficients are given
in terms of the initial datum by
\be a_\ell = {{\ell !}\over{\Gamma(c+\ell)}}\int_0^\infty
A(x,0)L_\ell^{c-1}(x)\,\,e^{-x}x^{c-1}dx.
    \lb{CC5} \ee
We assume the standard normalization of the Laguerre polynomials, \cite{Szego},
5.1.1.

An important conclusion follows immediately from the fact that the expansion
coefficients may
also be obtained from the solution $A(x,\tau)$:
\be a_\ell e^{-\ell\tau} = {{\ell !}\over{\Gamma(c+\ell)}}\int_0^\infty
A(x,\tau)
L_\ell^{c-1}(x)\,\,e^{-x}x^{c-1}dx. \lb{CC6} \ee
If we recall that $\Gamma(x,\tau)=A(x,\tau)e^{-x}$ and that $e^\tau=
L^{\gamma}(t)$, then (\ref{CC5})
and (\ref{CC6}) together imply that
\be [L(t)]^{\ell\gamma} \int_0^\infty
\Gamma(x,\tau)L_\ell^{c-1}(x)\,\,x^{c-1}dx
        =\int_0^\infty \Gamma(x,0)L_\ell^{c-1}(x)\,\,x^{c-1}dx. \lb{CC7} \ee
In other words, the lefthand side of this equation is for each $\ell=0,1,2,...$
an invariant
of motion of the equation (\ref{C4}) (with $a=c$). If we return to the unscaled
solution, these
invariants, with an appropriate choice of normalization, take the form
\be J_\ell(t):=  \int_{{\Bbb R}^d}L^{\ell\gamma}(t){{\ell !}\over{(c)_\ell}}
 L^{c-1}_\ell\left({{r^\gamma}\over{\gamma
L^\gamma(t)}}\right)\Theta(r,t)\,\,d^d\br \lb{C5n} \ee
for $\ell=0,1,2,...$. {\it Thus, there is an infinite sequence of integral
invariants of the equation
(\ref{1.3}) for} $\Theta(r,t)$. The first such invariant for $\ell=0$ is
nothing but the
Corrsin invariant $J_0=K$. The next two are, for $\ell=1$,
\be J_1(t):= \int_{{\Bbb R}^d}\left[ L^\gamma(t)-{{r^\gamma}\over{d}}\right]
                                                      \Theta(r,t)\,\,d^d\br
\lb{C5l} \ee
and for $\ell=2$,
\be J_2(t):= \int_{{\Bbb R}^d}\left[
L^{2\gamma}(t)-{{2L^\gamma(t)r^\gamma}\over{d}}
+{{r^{2\gamma}}\over{d(d+\gamma)}}\right]\Theta(r,t)\,\,d^d\br. \lb{C5m} \ee
These are really ``generalized invariants'', because they depend not only upon
the solution
$\Theta(r,t)$ but also explicitly upon the time $t$. However, we see that, in
the subspace defined
by the vanishing of the first $p$ invariants, $J_0=J_1=\cdots=J_{p-1}=0$, the
$p$th integral
$J_p\propto\int r^{p\gamma}\Theta(r,t)\,\,d^d\br$ and is an ordinary invariant.
This may also be inferred
directly from equation (\ref{1.3}), by using the fact that its righthand side
defines
an operator homogeneous of degree $-\gamma$ and using integration by parts.

These new invariants play a key role in the problem of the convergence in the
scaling limit.
We see using the relation $\Gamma(x,\tau)=A(x,\tau)e^{-x}$, the definition of
$\Phi_\ell(x)$,
and the expansion formulae (\ref{CC4}),(\ref{CC5}) that
\be \Gamma(x,\tau) = \sum_{\ell=0}^\infty c_\ell
[L(t)]^{-\gamma\ell}\Phi_\ell(x) \lb{CC8} \ee
with convergence of the summation in the $L^2$-sense on $(0,\infty)$ for the
weight $\tilde{w}_c(x)
= e^x x^{c-1}$. The expansion coefficient
\be c_\ell = {{1}\over{\Gamma(c)}}\int_0^\infty
\Gamma(x,0)L_\ell^{c-1}(x)x^{c-1}dx \lb{CC9} \ee
is proportional to the invariant $J_\ell,\,\,\ell=0,1,2,...$. Now suppose that
the first $p$
invariants vanish: $J_0=J_1=\cdots=J_{p-1}=0$. Since the solution $\Phi_p(x)$
corresponds
to $\alpha=d+p\gamma$, we see that if we scale according to that $\alpha$, we
obtain
\be \Gamma(x,\tau) = \sum_{\ell=0}^\infty
c_{p+\ell}[L(t)]^{-\gamma\ell}\Phi_{p+\ell}(x).
    \lb{CC10} \ee
Hence, it follows that $\lim_{\tau\rightarrow\infty}\Gamma(x,\tau)= c_p
\Phi_p(x)$.
Gathering together the above results, we may state the following proposition:
{\it Suppose that
the initial datum $\Theta(r,0)$ lies in the $L^2$ space with stretched
exponential weight
$e^{{{r^\gamma}\over{\gamma}}}$ and that the first $p$ invariants vanish
$J_0=J_1=\cdots=J_{p-1}=0$
but $J_p\neq 0$. If one scales the solution $\Theta(r,t)$ with
$\alpha=d+p\gamma$, then
\be \lim_{\tau\rightarrow\infty}\Gamma_{(\alpha)}(x,\tau) = c_p \Phi_p(x)
\lb{CC11} \ee
with convergence in the $L^2$ sense on $(0,\infty)$ with weight
$\tilde{w}_c(x)=e^x x^{c-1}.$
The constant $c_p$ is given by the ratio of the $p$th invariants $J_p$ of the
initial datum $\Gamma(x,0)$
and $\tilde{J}_p$ of the equilibrium solution $\Phi_p(x)$.} It only remains to
justify the last claim.
In fact, with the normalization of the invariants adopted here
\begin{eqnarray}
\tilde{J}_p  & = & {{1}\over{\Gamma(c)}}\int_0^\infty
\Phi_p(x)L_p^{c-1}(x)x^{c-1}dx \cr
          \, & = & {{p!}\over{\Gamma(p+c)}}\int_0^\infty
\left[L_p^{c-1}(x)\right]^2 e^{-x}x^{c-1}dx \cr
          \, & = & 1. \lb{M1}
\end{eqnarray}
This completes the proof.

It is interesting to note that $J_0=J_1=0$ but $0<J_p<\infty$ for some $p\geq
2$ is not consistent
with realizable initial data, due to the nonrealizability of scaling solutions
$\Phi_p(x)$
for $p>1$. Indeed, if $p$ is the least integer $p$ for which $J_p\neq 0$,
then our preceding result implies {\it a fortiori} that
$\Gamma(x,\tau)\rightarrow c_p \Phi_p(x)$
in $L^2$ with respect to the finite measure $e^{-x}x^{c-1}dx$ and hence, along
a subsequence of times
$\tau_k,\,k=1,2,...,$ convergence for a.e. {\it x.} If the initial data were
realizable (positive-definite),
then, since positive-definiteness is preserved by the dynamics and by pointwise
limits, the limit
$c_p\Phi_p(x)$ would be positive-definite as well. However, this
contradicts our earlier result. Thus, no positive-definite
$\Theta(r,0)$ can have $J_0=J_1=0$ but $0<J_p<\infty$ for some $p\geq 2.$ Our
argument here is rather
indirect, using the equation (\ref{1.3}), but the conclusion involves
no dynamics. In fact, it follows directly by the same argument used in
section 3.3 to prove nonrealizability of the scaling solutions for
$\nu>\gamma$ that any initial datum $\Theta(0)\in L^2$ (unweighted) with
$J_0=J_1=0$
cannot be realizable (positive-definite). The condition
$J_0=J_1=0$ is precisely equivalent to the condition of vanishing moments,
(\ref{add1}), employed there.

\newpage

There is a direct connection (pointed out to us by K. Gaw\c{e}dzki) of our
generalized invariants with the
``slow modes'' in 2-particle Lagrangian statistics that were discovered in
\cite{BGK}. It is a consequence of that work
that there are homogeneous moment functions $\phi_{0,p}(\br)$ of degree
$\sigma_{0,p},\,\,p=0,1,2,...$, whose integrals
over any initial 2-point function evolve in time as pure degree $p$
polynomials:
\be  \int \phi_{0,p}(\br)\,\,\Theta(\br,t)\,\,d^d\br
     = \sum\limits_{q=0}^p c_{p,q}\,t^{p-q}\int\phi_{0,q}(\br)\,
\Theta(\br,0)\,\, d^d\br, \lb{add7} \ee
for some computable constants
$c_{p,q}=\int\overline{\psi_{0,q}(\br,1)}\,\phi_{0,p}(\br) d^d\br,$ in the
notations
of \cite{BGK}. As these constants are manifestly independent of initial data,
it is not hard to infer from (\ref{add7})
the existence of an associated sequence of ``generalized invariants''. These
``slow modes'' were constructed
in \cite{BGK} for every angular momentum sector $\ell=0,1,2,...$, but for the
rotationally-invariant sector $\ell=0$
they are particularly simple, given just by the powers
$\phi_{0,p}(r)=r^{p\gamma}$. See Appendix A of \cite{BGK}. It can then be
easily shown
that the associated sequence of generalized invariants in the $\ell=0$ sector
coincides with the sequence $J_p,\,\,p=0,1,2,...$ we found above.

\noindent {\it (4.3) Initial Data with Slow Decay \& A Finite Invariant}

\noindent The previous results do not allow us to address the question whether
permanence of large eddies
(PLE) holds in the space domain. For this, we must consider initial data with
only power-law decay
at large distances. Such data with slow decay fall themselves into two broad
classes: those
with one of the invariants $J_\ell,\,\,\ell=0,1,2,...$ finite and those with no
finite
invariants. Here by ``finite'' we mean both non-zero and non-infinite. It will
be shown below
that the class of initial data with a finite invariant behaves very
similarly---as far as the
leading-order behavior is concerned---to the initial data with rapid  decay.

To study such initial data, a new technique is required. An important
observation
is that (\ref{C5}) is the backward equation corresponding to the {\it
Fokker-Planck equation}
on the half-line $x>0$
\be {{\partial P}\over{\partial\tau}}(x,\tau)
     = -{{\partial}\over{\partial x}}\left(a(x)P(x,\tau)\right)
 + {{1}\over{2}}{{\partial^2}\over{\partial x^2}}\left(b(x)P(x,\tau)\right)
\label{C5a} \ee
with drift
\be    a(x)= c+ x + 2 x {{\Phi'(x)}\over{\Phi(x)}} \label{C5b} \ee
and diffusion
\be    b(x)= \sigma^2(x)= x. \lb{C5c} \ee
Thus, it follows immediately that
\be    A(x,\tau)= E[A(X_{x,\tau},0)]=\int_0^\infty dy\,\,A(y,0)P(y,\tau|x,0)
\lb{C5d} \ee
where $X_{x,\tau}$ is the diffusion starting at $x$ at $\tau=0$ and obeying the
Ito equation
\be       dX_\tau = a(X_\tau)d\tau + \sigma(X_\tau)dW_\tau, \lb{C5e} \ee
and $P(y,\tau|x,0)$ is the transition probability kernel for the process. The
formal invariant
density $P(x)$  of the Fokker-Planck equation is
\be    P(x)\propto x^{c-1} e^x \Phi^2(x). \lb{C5f} \ee
However, this is only a normalizable probability density for the exceptional
cases when
$\nu=\gamma\ell,\,\,\ell=0,1,2,...$, in which case we refer to the density as
$P_\ell(x),\,\,\ell=0,1,2,...$. This can be easily understood from the
character of the
drift term. When $\Phi(x)\propto x^{-a}$ at large $x$, then $a(x)\sim c-2a + x$
for $x\gg 1$, which
is unstable. However, when $\Phi(x)=p(x)e^{-x}$ for some polynomial $p(x)$---as
for the exceptional
series---then $a(x)\sim a-x$ for $x\gg 1$, which is
stable. \footnote{These observations give another method to infer the
  existence of the sequence of generalized invariants
  $J_p,\,\,p=0,1,2,...$ In fact, using the invariant density in
  (\ref{C5f}), equation (\ref{C5})  may be rewritten as
$$ {{\partial A}\over{\partial\tau}}(x,\tau)= {{x^{1-c}
    e^{-x}}\over{\Phi^2(x)}}{{\partial}\over{\partial x}}
   \left[x^c e^x\Phi^2(x){{\partial A}\over{\partial x}}(x,\tau)\right]. $$
It then follows by a formal integration by parts argument that
$$  J(\tau) := \int_0^\infty dx\,\,x^{c-1} e^x A(x,\tau) \Phi^2(x) $$
is an invariant for {\it all} $\alpha$. However, only for the
exceptional values $\alpha= d+\gamma p,\,\,p=0,1,2,...$ do the
integrands decay rapidly enough to justify this argument. These give
the familiar sequence of invariants $J_p,\,\,p=0,1,2,...$}

In particular, it is true for the equipartition solution $\Phi_0(x)=e^{-x}$
that
\be a(x)= c-x \lb{C5f0} \ee
and the invariant distribution is given by
\be P_0(x)={{1}\over{\Gamma(c)}}x^{c-1} e^{-x}. \lb{C5f1} \ee
For this $\ell=0$ process we can also calculate the transition probability
kernel by means of expansion
in the eigenfunctions of the backward operator ${\cal L}_0$ in section 4.1:
\be P(y,\tau|x,0)= {{y^{c-1} e^{-y}}\over{\Gamma(c)}}
  \sum_{\ell=0}^\infty {{L_\ell^{c-1}(y)L_\ell^{c-1}(x)}
\over{\left({\,\,\,\,\,\,}_\ell^{\!\!\!\!\!\!\!\ell+c-1}\right)}}e^{-\ell\tau}.
\lb{C5f2} \ee
This sum can be evaluated in closed form, using \cite{Erd}, 10.12(20), with
$\alpha=c-1,
z=e^{-\tau}$:
\be P(y,\tau|x,0)= {{y^{c-1}
e^{-y}}\over{1-e^{-\tau}}}\exp\left\{-{{x+y}\over{e^\tau-1}}\right\}
    \left(xye^{-\tau}\right)^{-(c-1)/2}
I_{c-1}\left\{{{2\left(xye^{-\tau}\right)^{1/2}}\over{1-e^{-\tau}}}\right\},
\lb{C5f3} \ee
in terms of the modified Bessel function $I_{c-1}(z)$. Cf. \cite{CFKL},
equation (2.15b).

Using these results, we can establish our first main convergence result of this
section:
{\it If $\Theta(0)\in L^\infty\bigcap L^1$ and the Corrsin invariant of the
initial datum is non-zero,
then the limit exists,}
$$ \lim_{\tau\rightarrow\infty}\Gamma_{(d)}(x,\tau)= c_0 \Phi_0(x) $$
{\it uniformly on compacts in $x$. The constant $c_0$ is the ratio of the
Corrsin invariants $K$
of the initial data $\Gamma(x,0)$ and $\tilde{K}$ of the equipartition scaling
solution $\Phi_0(x)$.}
The condition $\Theta(0)\in L^\infty$ is natural, since $|\Theta(r,0)|\leq
\Theta(0,0)<\infty$ for
any positive-definite initial data with finite energy.
This theorem includes the case of power-law decay $\Theta(r,0) \sim A
r^{-\alpha}$ for $r\gg L_0$
with $\alpha>d$, which guarantees integrability. Because $\Theta(0)\in L^1$,
the Corrsin invariant
must be finite. The theorem remains true even if the Corrsin invariant is zero
initially, but in that
case it yields a trivial (null) scaling limit.

We prove the result first for bounded $A(0)$. This will follow from standard
convergence results
for time-dependent distributions of one-dimensional diffusion processes. For
example, \cite{GS},
Section 23, Theorem 3 implies that for any $A(0)\in L^\infty$ and uniformly on
compacts in $x$,
\be \lim_{\tau\rightarrow\infty}E\left[A\left(X_{x,\tau},0\right)\right]=\int
A(y,0)\,P(y)\,dy
     \lb{CX1} \ee
where
\be P(x)\propto {{1}\over{\sigma^2(x)}}\exp\left\{\int_{x_0}^x
{{2a(t)}\over{\sigma^2(t)}}dt\right\}.
     \lb{CX2} \ee
is the invariant measure of the process, if the diffusion is defined with
instantaneous
reflection upon hitting the boundaries. Such boundary conditions must therefore
be checked to be
satisfied. We show, in fact, that the process $X_{x,\tau}$ has zero probability
of reaching either
of the boundaries of the semi-infinite interval, $0$ or $\infty$, at any finite
time. Thus, it is a
special case of reflected b.c., with no reflection ever required, and the
Theorem 3 of \cite{GS} applies.

We will treat the boundary conditions in a generality that will permit to
discuss later cases as well.
According to \cite{GS}, Section 21, Theorem 1, for any $b\in (0,\infty)$, if
\be  L_1^-=\int_0^b\exp\left\{ -\int_b^x {{2a(t)}\over{\sigma^2(t)}}dt\right\}
dx= +\infty, \lb{C5f4} \ee
then the process $X_{x,\tau}$ attains the point $b$ before $0$ a.s., for any
$x\in (0,b)$.
In that case, the process never hits $0$ a.s., because continuity in time
requires that it
pass through the interval $(0,b)$ to reach $0$ and the Markov property requires
that each time
it re-enters the interval it must exit through $b$. The same statement holds
for the right endpoint: if
\be  L_1^+=\int_b^\infty
          \exp\left\{ -\int_b^x {{2a(t)}\over{\sigma^2(t)}}dt\right\} dx=
+\infty, \lb{C5f4a} \ee
then the process $X_{x,\tau}$ attains the point $b$ before reaching $\infty$
a.s., for any
$x\in (b,\infty)$, and thus never reaches $\infty$ in any finite time a.s. We
note in general that
\be \psi(x;b):=\exp\left\{-\int_{b}^x {{2a(t)}\over{\sigma^2(t)}}dt\right\}=
               C\left[\sigma^2(x)P(x)\right]^{-1}, \lb{Cx3} \ee
using (\ref{CX2}), where the constant $C=\sigma^2(b)P(b)$. Thus, for any of the
processes (\ref{C5a}),
(\ref{C5b}), it follows that
\be \psi(x;b)= C' {{e^{-x}}\over{x^c\Phi^2(x)}} \lb{Cx4} \ee
using $\sigma^2(x)=2x$ and the formula (\ref{C5f}) for the invariant measure.
Clearly,
$L_1^-=\int_0^b \psi(x;b)\,\,dx$ and $L_1^+=\int_b^\infty \psi(x;b)\,\,dx$.

In the case at hand, for the equipartition solution, we see that $\psi_0(x;b)=C
x^{-c}e^x$.
Since $c\geq 1$, both $L_1^-=+\infty$ and $L_1^+=+\infty$. Thus, we conclude
that the
process never reaches the boundary in finite time a.s. This completes the
convergence
proof for the case of bounded $A(0)$. By itself, this is only a strengthening
of the result in section 4.1, since $A(0)\in L^\infty$ implies that $A(0)$ is
$L^2$ with respect
to the weight $w_c(x)=x^{c-1}e^{-x}$. However, the sense of convergence is
stronger,
being now pointwise in $x$ uniformly on compacts.

Having proved the result for bounded $A(0)$, we now extend to the case where
$\Gamma(0)\in L^\infty$
and $\Gamma(0)\in L^1$ with respect to weight $x^{c-1}$. We have, using
$\Gamma(x,0)=A(x,0)e^{-x},$ that
\be  |A(x,0)|\leq \|\Gamma(0)\|_{L^\infty} e^x \lb{C5g0} \ee
and
\be \int_0^\infty |A(x,0)| P_0(x)\,\,dx
    = {{1}\over{\Gamma(c)}}\int_0^\infty |\Gamma(x,0)|\,\,x^{c-1}dx<\infty.
\lb{C5g1} \ee
Because of the latter result, we may choose $M>1$ so large that for any small
$\epsilon>0$,
\be \int_M^\infty |A(x,0)| P_0(x)\,\,dx<\epsilon. \lb{C5g2} \ee
Let us then take
\be A^{(M)}(x,0):= \left\{ \begin{array}{ll}
                           A(x,0) & x\leq M \cr
                             0    & x> M
                           \end{array} \right. \lb{C5g3} \ee
Thus, by (\ref{C5g0}), $A^{(M)}(0)$ is bounded: $\|A^{(M)}(0)\|_{L^\infty}\leq
\|\Gamma(0)\|_{L^\infty} e^M.$
By triangle inequality
\begin{eqnarray}
 \, & & \left|\int_0^\infty dy\,\,A(y,0)P(y,\tau|x,0)-\int_0^\infty
dy\,\,A(y,0)P_0(y)\right| \cr
 \,\,\,\,\,\,\,\,\,\,\,\,\,\,\,\,\,\,\,\, & & \,\,\,\,\,\,\,\,\,\,
                                              \leq \,\,\,\,\left|\int_0^\infty
dy\,\,A^{(M)}(y,0)
                                               P(y,\tau|x,0)-\int_0^\infty
dy\,\,
                                               A^{(M)}(y,0)P_0(y)\right| \cr
\,& &\,\,\,\,\,\,\,\,\,\,\,\,\,\,\,\,\,\,\,\,\,\,\,\,\,\,\,\,\,\,
     + \int_M^\infty dy\,\,|A(y,0)| P(y,\tau|x,0) + \int_M^\infty
dy\,\,|A(y,0)| P_0(y). \lb{C5g4}
\end{eqnarray}

To control the second term we employ the estimate
\be \sup_{y>0} {{P(y,\tau|x,0)}\over{P_0(y)}} \leq
{{e^x}\over{(1-e^{-\tau})^c}}. \lb{C5g5} \ee
This is proved using the inequality (\cite{Erd}, 7.3.2(4) and 7.2.2(12)):
\be |z^{-\nu}I_\nu(z)|\leq {{e^{|{\rm Re}\,z|}}\over{2^\nu \Gamma(\nu+1)}}
\lb{C5g6} \ee
with $z={{2\left(xye^{-\tau}\right)^{1/2}}\over{1-e^{-\tau}}}$ and $\nu=c-1$ in
(\ref{C5f3}). Thus,
\be P(y,\tau|x,0) \leq {{P_0(y)}\over{(1-e^{-\tau})^c}}
\exp\left\{{{-(x+y)+2\left(xye^{\tau}\right)^{1/2}}\over{e^\tau-1}}\right\}.
\lb{C5g7} \ee
The maximum of the exponent is found to occur at $y=xe^\tau$, yielding the
estimate (\ref{C5g5}).
That inequality can then be used to bound
\be \int_M^\infty dy\,\,|A(y,0)| P(y,\tau|x,0) \leq
{{e^x}\over{(1-e^{-\tau})^c}}\epsilon \lb{C5g8} \ee
because of the condition (\ref{C5g2}) on $M$. Since $A^{(M)}(0)$ is bounded for
a fixed $M$,
our preliminary convergence result applies and we obtain
\be \limsup_{\tau\rightarrow\infty}
    \left|\int_0^\infty dy\,\,A(y,0)P(y,\tau|x,0)-\int_0^\infty
dy\,\,A(y,0)P_0(y)\right|
    \leq (e^x+1)\epsilon. \lb{C5g9} \ee
Since $\epsilon$ was arbitrary, we get
\be \lim_{\tau\rightarrow\infty}\int_0^\infty dy\,\,A(y,0)P(y,\tau|x,0)
     =\int_0^\infty dy\,\,A(y,0)P_0(y):=c_0. \lb{C5g10} \ee
and therefore
\be \lim_{\tau\rightarrow\infty}\Gamma(x,\tau)= c_0 \Phi_0(x) \lb{C5i} \ee
uniformly on compact subsets of $(0,\infty)$.

To identify the constant $c_0$, we note that the Corrsin invariant of the
initial data is
\be K(0)= \gamma^{c-1}\omega_{d-1}\int_0^\infty
A(x,0)\Phi_0(x)\,\,x^{c-1}\,\,dx. \lb{C5j} \ee
Furthermore, $P_0(x)=
{{1}\over{\widetilde{K}}}\gamma^{c-1}\omega_{d-1}x^{c-1}\Phi_0(x),$
where $\widetilde{K}$ is the Corrsin invariant of the equipartition solution
$\Phi_0$. Thus,
\be K(0)=\widetilde{K}\cdot \int_0^\infty A(x,0)P_0(x)\,\,dx =
\widetilde{K}\cdot c_0.
    \lb{C5k} \ee
Of course, this is consistent with the time-invariance of the Corrsin integral,
which,
by (\ref{C5i}), gives $K(\tau)= c_0\widetilde{K}$ for all $\tau\geq 0$.

If the Corrsin invariant is finite but vanishes, then the results in the
preceding Section 4.2
suggest that the asymptotic behavior will be described by $\Phi_p$, if $J_p$ is
the first
non-vanishing invariant. We next prove such a result, which is relevant to
initial conditions
with power-law decay for $\nu>\gamma$. The theorem we establish is the
following:
{\it Suppose that $\Theta(0)\in L^\infty$ and $\int_{{\Bbb R}^d}
d\br\,\,r^{p\gamma}|\Theta(r,0)|
<\infty.$ If $J_\ell=0,\,\,\ell=0,1,...,p-1$ but $J_p\neq 0,$ then the solution
rescaled appropriate
to parameter $\alpha=d+p\gamma$ satisfies
\be    \lim_{\tau\rightarrow\infty}\Gamma_{(\alpha)}(x,\tau)=c_p \Phi_p(x)
\lb{C5r} \ee
uniformly on compact subsets of $x$. Furthermore, the constant $c_p$ is the
ratio $J_p/\widetilde{J}_p,$
where $\widetilde{J}_p$ is the value of the invariant for the scaling solution
$\Theta_p(r,t)=
\vartheta^2(t)\Phi_p(r/L(t)).$}

We give the proof first for the bounded $A(0)$. We wish to make a proof very
similar to the previous one.
In fact,
\be     A(x,\tau)= E[ A(X_{x,\tau},0)] \lb{C5s} \ee
where $X_{x,\tau}$ is the diffusion process appropriate to $\alpha=d+p\gamma.$
It is enough to show that
\be \lim_{\tau\rightarrow\infty}A(x,\tau)=\int_0^\infty A(x,0)P_p(x)\,\,dx:=c_p
\lb{C5t} \ee
uniformly on compact sets of $x$ with
\be P_p(x) = {{p!}\over{\Gamma(c+p)}}\left[L_p^{c-1}(x)\right]^2 e^{-x}x^{c-1}
\lb{C5t1} \ee
the invariant measure of the process. The complication is that the diffusion in
all the cases
$p\geq 1$ has singular points and decomposes into $p+1$ simple pieces, each
with its own
invariant measure. In fact, using (\ref{C5a}) for the drift $a(x)$ and
(\ref{3.29}) for $\Phi_p(x)$,
\be a(x)= c-x +\sum_{k=1}^p {{2x}\over{x-x_k}} \lb{C5t2} \ee
where $x_k,\,\,k=1,...,p$ are the $p$ roots of the generalized Laguerre
polynomial $L_p^{c-1}(x)$.
We recall that these are all real and simple, and located in the interior of
the interval $(0,\infty)$.
Cf. \cite{Szego}, Theorem 3.3.1. We may label them in increasing order
$x_1<x_2<\cdots<x_p$.
Because of the pole terms, the zeros are repulsive singularities of the drift
field $a(x)$ and the regular
set of points of the process decomposes into a disjoint union of $p+1$ open
intervals, $I_k=(x_k,x_{k+1}),
\,\,k=0,1,..,p,$ with $x_0=0$ and $x_{p+1}=\infty$. It is easy to calculate
that for each $x,b\in I_k$,
\be \psi(x;b)= {{C x^{-c}e^{x}}\over{\prod_{k=1}^p (x-x_k)^2}}, \lb{C5t3} \ee
with $C$ some constant. Thus, $\int_{x_k}^b \psi(x;b)\,\,dx=+\infty$ and
$\int_b^{x_{k+1}}
\psi(x;b)\,\,dx=+\infty$, so that, again by \cite{GS}, Section 21, Theorem 1,
the boundary points
of each interval are inaccessible in finite time a.s. Thus, the process is not
ergodic but
instead there exist $p+1$ distinct, ergodic invariant distributions $P_p^{(k)}$
supported on the
intervals $I_k,\,\,k=0,1,...,p.$ Up to a normalization factor
$w_k=(\int_{I_k}P_p(x)dx)^{-1}$,
these coincide with $P_p$ restricted to the interval $I_k$, i.e.
\be P_p^{(k)}(x) = \left. w_k P_p(x)\right|_{I_k} \lb{C5t4} \ee
On each interval separately, the Theorem 3, Section 23 of \cite{GS} applies.
Thus, for $x\in I_k$
\be  \lim_{\tau\rightarrow\infty}E\left[A(X_{x,\tau},0)\right]= w_k \int_{I_k}
A(x,0)P_p(x):=c_{p,k}.
     \lb{C5t5} \ee
We need to show that the constants $c_{p,k}$ are, in fact, independent of $k$.

By assumption, the initial data has the first $p$ invariants vanishing,
$J_0=\cdots=J_{p-1}=0$
but $J_p\neq 0$:
\be {{p!}\over{\Gamma(p+c)}}\int_0^\infty
A(x,0)L_\ell^{c-1}(x)L_p^{c-1}(x)e^{-x}x^{c-1}\,dx = 0
     \lb{C5t6} \ee
for $\ell=0,...,p-1$, and
\be {{p!}\over{\Gamma(p+c)}}\int_0^\infty
A(x,0)\left[L_p^{c-1}(x)\right]^2e^{-x}x^{c-1}\,dx = J_p.
     \lb{C5t7} \ee
The uniform bound $|A(x,\tau)|\leq \|A(0)\|_{L^\infty}$ follows from
(\ref{C5s}). Thus, the $J_\ell,
\,\,\ell=0,1,...p$ are rigorously dynamical invariants and the equations
(\ref{C5t6}),(\ref{C5t7})
hold with $A(x,0)$ replaced by $A(x,\tau)$. Then we may apply (\ref{C5t5}) and
Lebesgue's theorem
to conclude
\be {{p!}\over{\Gamma(p+c)}}\sum_{k=0}^p c_{p,k}\int_{x_k}^{x_{k+1}}
                            L_\ell^{c-1}(x)L_p^{c-1}(x)e^{-x}x^{c-1}\,dx = 0
\lb{C5t8} \ee
for $\ell=0,...,p-1$, and
\be {{p!}\over{\Gamma(p+c)}}\sum_{k=0}^p c_{p,k}\int_{x_k}^{x_{k+1}}
                            \left[L_p^{c-1}(x)\right]^2e^{-x}x^{c-1}\,dx = J_p.
\lb{C5t9} \ee
These may be summarized as a matrix equation ${\bf Mc}_p={\bf J}$ with
\be M_{\ell,k} := \int_{I_k} L_\ell^{c-1}(x)L_p^{c-1}(x)e^{-x}x^{c-1}\,dx.
\lb{C5t10} \ee
By orthogonality of Laguerre polynomials one solution is
$c_{p,k}=J_p/\tilde{J}_p$ for all $k$, where recall
\begin{eqnarray}
\tilde{J}_p & := &
            {{1}\over{\Gamma(c)}}\int_0^\infty \Phi_p(x)L_p^{c-1}(x)
x^{c-1}\,dx \cr
          \,& =  & {{p!}\over{\Gamma(p+c)}}\int_0^\infty
\left[L_p^{c-1}(x)\right]^2e^{-x}x^{c-1}\,dx =1
\lb{C5t11}
\end{eqnarray}
is the $p$th invariant of the scaling solution $\Phi_p$ itself. This is the
unique solution if
the matrix ${\bf M}$ is nonsingular. That is equivalent to the statement that
an arbitrary
polynomial $p(x)$ of degree $p$ can satisfy
\be  \int_{I_k} p(x)L_p^{c-1}(x) x^{c-1}\,dx=0,\,\,\,\,k=0,1,...,p \lb{C5t12}
\ee
only if $p(x)\equiv 0$. In fact, this is true, because the polynomial has at
most $p$
real roots but there are $p+1$ intervals. Hence, there is at least one interval
$I_k,\,\,k=0,1,...,p$
on which it does not change sign. Then (\ref{C5t12}) implies that $p(x)\equiv
0$ on that interval,
and, hence, everywhere. Thus, we conclude. As a by-product of this argument, we
have shown that
\be c_p= \int_0^\infty A(x,0)P_p(x)\,\,dx
          = {{1}\over{\Gamma(c)}}\int_0^\infty
\Gamma(x,0)L^{c-1}_p(x)\,\,x^{c-1}dx \lb{C5u} \ee
is given by $c_p=J_p/\widetilde{J}_p$, as claimed.

Having proved the result for bounded $A(0)$, we now extend to the case where
$\Gamma(0)\in L^\infty$
and $\Gamma(0)\in L^1$ with respect to weight $x^{p+c-1}$. Here we make use of
the observation
that the scalings by $\alpha_p=d+p\gamma$ and $\alpha_0=d$ are simply related
by
$\Gamma_{(\alpha_p)}(x,\tau)= e^{p\tau}\Gamma_{(d)}(x,\tau)$, where we have
made the $\alpha$-dependence
explicit. After this we shall employ the equipartition scaling and the
corresponding definition
of $\Gamma_{(d)}(x,\tau)=A_{(d)}(x,t)e^{-x}$, so, when no $\alpha$-dependence
is given, $\alpha=d$
is implied. Then, we must show that
\be \lim_{\tau\rightarrow\infty} e^{p\tau}A(x,\tau)= c_p
{{p!}\over{(c)_p}}L_p^{c-1}(x) \lb{C5u1} \ee
uniformly on compacts, where
\be c_p={{1}\over{\Gamma(c)}}\int_0^\infty
\Gamma(x,0)L^{c-1}_p(x)\,\,x^{c-1}dx. \lb{C5u2} \ee
We define the following auxilliary function:
\be H(y,0):= (-1)^p \int_y^\infty dy_1\int_{y_1}^\infty dy_2 \cdots
\int_{y_{p-1}}^\infty dy_p \,\,y_p^{c-1}
                                           \Gamma(y_p,0). \lb{C5u6} \ee
It is not hard to see that reversing orders of integrations by the Tonelli
theorem gives
\begin{eqnarray}
\int_0^\infty |H(y,0)|\,\,dy & \leq &
         \int_0^\infty dy_p \int_0^{y_p} dy_{p-1}\cdots \int_0^{y_1} dy
\,\,y_p^{c-1}|\Gamma(y_p,0)| \cr
       \, & = & {{1}\over{p!}}\int_0^\infty |\Gamma(y,0)|
\,\,y^{p+c-1}dy<\infty.
    \lb{C5u7}
\end{eqnarray}
Of course, it follows directly from the definition that
\be \Gamma(y,0)= {{1}\over{y^{c-1}}} {{d^p}\over{dy^p}}H(y,0). \lb{C5u8} \ee
Because of the vanishing of the first $p$ invariants, one may readily check
that
\be \left.{{d^\ell
H}\over{dy^\ell}}(y,0)\right|_{y=0}=0,\,\,\,\,\ell=0,1,...,p-1. \lb{C5u9} \ee
Thus, making $p$ integrations by parts with ${{d^p}\over{dy^p}}L_p^{c-1}(y)=
(-1)^p$
\begin{eqnarray}
{{1}\over{\Gamma(c)}}\int_0^\infty H(y,0)\,\,dy & = &
                     {{1}\over{\Gamma(c)}}\int_0^\infty
{{d^p}\over{dy^p}}H(y,0) L_p^{c-1}(y)\,\,dy \cr
                 \, & = & {{1}\over{\Gamma(c)}}\int_0^\infty
\Gamma(y,0)L_p^{c-1}(y)y^{c-1}dy =  c_p.
 \lb{C5u10}
\end{eqnarray}
Since the integral is absolutely convergent, we may choose $M$ sufficiently
large that
\be {{1}\over{\Gamma(c)}}\int_M^\infty |H(y,0)|\,\,dy<\epsilon. \lb{C5u3} \ee
Let us define a decomposition $H(y,0)=H^{(M)}(y,0)+\overline{H}^{(M)}(y,0)$ via
$H^{(M)}(y,0)=\varphi^{(M)}(y)H(y,0)$ and $\overline{H}^{(M)}(y,0)
=\overline{\varphi}^{(M)}(y)H(y,0)$ for
a smooth decomposition of unity
$\varphi^{(M)}(y)+\overline{\varphi}^{(M)}(y)=1,$ with
$\varphi^{(M)}(y),\overline{\varphi}^{(M)}(y)\geq 0$ and
$\varphi^{(M)}(y)=0$ for $y>M+1$, $\overline{\varphi}^{(M)}(y)=0$ for $0<y<M$.
We may then define a corresponding decomposition
$\Gamma(y,0)=\Gamma^{(M)}(y,0)+\overline{\Gamma}^{(M)}(y,0)$
via (\ref{C5u8}), and likewise for
$A(y,0)=A^{(M)}(y,0)+\overline{A}^{(M)}(y,0)$ and
$c_p=c_p^{(M)}+\overline{c}_p^{(M)}$.
%
%
Next we can employ the transition probability of the $\alpha=d$ process in a
triangle inequality:
\begin{eqnarray}
 \, & & \left|e^{p\tau}\int_0^\infty dy\,\,A(y,0)P(y,\tau|x,0)-
         {{p!}\over{\Gamma(c+p)}}L_p^{c-1}(x)\int_0^\infty dy\,\,H(y,0)\,
\right| \cr
 \,\,\,\,\,\,\,\,\,\,\,\,\,\,\,\,\,\,\,\, & & \,\,\,\,\,\,\,\,\,\,
         \leq \,\,\,\,\left|e^{p\tau}\int_0^\infty
dy\,\,A^{(M)}(y,0)P(y,\tau|x,0)
                         -c_p^{(M)}{{p!}\over{(c)_p}}L_p^{c-1}(x)\right| \cr
\,& &\,\,\,\,\,\,\,\,\,\,\,\,\,\,\,\,\,\,\,\,\,\,\,\,\,\,\,\,\,\,
     +  e^{p\tau}\left|\int_M^\infty
dy\,\,\overline{A}^{(M)}(y,0)P(y,\tau|x,0)\right| \cr
\,& &\,\,\,\,\,\,\,\,\,\,\,\,\,\,\,\,\,\,\,\,\,\,\,\,\,\,\,\,\,\,
     + {{p!}\over{\Gamma(c+p)}}|L_p^{c-1}(x)|\int_M^\infty
dy\,\,\left|\overline{H}^{(M)}(y,0)\right|.
       \lb{C5u5}
\end{eqnarray}
As before, $\|A^{(M)}(0)\|_{L^\infty}\leq \|\Gamma(0)\|_{L^\infty} e^{(M+1)}.$
Thus, we can identify the
first term as $\left|A_{\alpha_p}^{(M)}(x,\tau)-c_p^{(M)}\right|\cdot
{{p!}\over{(c)_p}}|L_p^{c-1}(x)|$
and appeal to our preliminary result for $L^\infty$ initial data to conclude
that this goes
to zero uniformly on compact sets of $x$. Of course, the third term is less
than
${{p!}\over{(c)_p}}|L_p^{c-1}(x)|\cdot\epsilon$ by the assumption (\ref{C5u3}).
The main problem is to control the middle term.

If we substitute the expression (\ref{C5u8}) for $\overline{\Gamma}^{(M)}(y,0)$
in terms
$\overline{H}^{(M)}(y,0)$ and integrate by parts $p$ times, we obtain
\be \int_M^\infty dy\,\,\overline{A}^{(M)}(y,0)P(y,\tau|x,0)=
                 \int_M^\infty dy\,\,
\overline{H}^{(M)}(y,0)\,\,\left(-{{d}\over{dy}}\right)^p
                 \left[{{P(y,\tau|x,0)}\over{y^{c-1}e^{-y}}}\right]. \lb{C5u11}
\ee
Employing the formula (\ref{C5f3}) for the transition probability gives
\be \int_M^\infty dy\,\,\overline{A}^{(M)}(y,0)P(y,\tau|x,0)=
     {{2^{c-1}}\over{(1-e^{-\tau})^c}}\int_M^\infty dy\,\,
\overline{H}^{(M)}(y,0)\,\,\left(-{{d}\over{dy}}\right)^p
     \left[\exp\left\{-{{x+y}\over{e^\tau-1}}\right\}
z^{-(c-1)}I_{c-1}(z)\right], \lb{C5u12} \ee
with $z={{2\left(xye^{-\tau}\right)^{1/2}}\over{1-e^{-\tau}}}$. The derivative
can be evaluated
by the generalized product rule $D^p(uv)=\sum_{r=0}^p \left({\,}_r^p\right) D^r
u\cdot D^{p-r}v$
and the relation
\be {{d}\over{dy}}= {{2x e^{-\tau}}\over{(1-e^{-\tau})^2}}{{d}\over{zdz}}.
\lb{C5u13} \ee
We note that
\be {{d^r}\over{dy^r}}\exp\left\{-{{x+y}\over{e^\tau-1}}\right\}
                 = (-1)^r (e^{\tau}-1)^{-r} \cdot
\exp\left\{-{{x+y}\over{e^\tau-1}}\right\}.
      \lb{C5u14} \ee
Likewise, defining $\xi= x/(1-e^{-\tau})$,
\be {{d^r}\over{dy^r}}\left[z^{-(c-1)}I_{c-1}(z)\right]
     = {{2^r \xi^r
}\over{(e^{\tau}-1)^{r}}}\left[z^{-(c+r-1)}I_{c+r-1}(z)\right] \lb{C5u15} \ee
using (\ref{C5u13}) and
$\left({{d}\over{zdz}}\right)^r\left[z^{-\nu}I_{\nu}(z)\right]=
z^{-(\nu+r)}I_{\nu+r}(z),$ \cite{Erd}, 7.11(20). Summing over all the
contributions
and using the estimate (\ref{C5g6}) for the Bessel function gives
\be \left|{{d^p}\over{dy^p}}\left[\exp\left\{-{{x+y}\over{e^\tau-1}}\right\}
    z^{-(c-1)}I_{c-1}(z)\right]\right|
    \leq {{p! {L}_p^{c-1}(-\xi)}\over{2^{c-1}\Gamma(c+p)(e^\tau-1)^p}}
\exp\left\{{{-(x+y)+2\left(xye^{\tau}\right)^{1/2}}\over{e^\tau-1}}\right\}.
\lb{C5u16} \ee
We employed the relation
\be L_p^{c-1}(x)=\sum_{r=0}^p
\left({\,}_{\,\,\,p-r}^{c+p-1}\right){{(-x)^r}\over{r!}} \lb{C5u66} \ee
in \cite{Erd}, 10.12(7). Thus, $|L_p(x)|\leq L_p^{c-1}(-x)$. The exponential
term is the same as was
shown before to be bounded over all $y$ by $e^{x}$. Thus, we obtain finally
from (\ref{C5u12}) that
\be \left| e^{p\tau}\int_M^\infty
dy\,\,\overline{A}^{(M)}(y,0)P(y,\tau|x,0)\right|\leq
     {{p!L_p^{c-1}(-\xi)e^x}\over{(1-e^{-\tau})^{c+p}}} \cdot
     {{1}\over{\Gamma(c+p)}}\int_M^\infty dy\,\, |H(y,0)| \lb{C5u17} \ee
Thus,
\be \limsup_{\tau\rightarrow\infty}\left| e^{-p\tau}\int_M^\infty
dy\,\,\overline{A}^{(M)}(y,0)P(y,\tau|x,0)\right|
           \leq {{p!}\over{(c)_p}}L_p^{c-1}(-x)e^{x}\cdot \epsilon, \lb{C5u18}
\ee
by the definition (\ref{C5u3}) of $M$. Since $A^{(M)}(0)$ is bounded for a
fixed $M$,
our preliminary convergence result applies and we obtain
\be \limsup_{\tau\rightarrow\infty}
    \left|e^{p\tau}\int_0^\infty dy\,\,A(y,0)P(y,\tau|x,0)-c_p
{{p!}\over{(c)_p}}L_p^{c-1}(x)\right|
    \leq (e^x+1){{p!}\over{(c)_p}}L_p^{c-1}(-x)\cdot \epsilon. \lb{C5u19} \ee
Since $\epsilon$ is arbitrary, we get
\be \lim_{\tau\rightarrow\infty} e^{p\tau} A(x,\tau) = c_p
{{p!}\over{(c)_p}}L_p^{c-1}(x) \lb{C5u20} \ee
uniformly on compact subsets of $(0,\infty)$ and likewise
\be \lim_{\tau\rightarrow\infty}\Gamma_{\alpha_p}(x,\tau)= c_p \Phi_p(x).
\lb{C5u21} \ee

The most surprising consequence, in view of traditional beliefs, is the implied
breakdown
of spatial PLE on length-scales comparable to the integral scale $L(t)$.
Naively one would
expect that PLE holds in space for all long-range powers laws, $\sim
Ar^{-\alpha}$ with any
$\alpha>0$. Instead, the memory of the initial conditions is through the
invariant $J_p$ and not
through the initial amplitude $A$ of the power-law tail, when
$\alpha>d+p\gamma$ and $J_p\neq 0$.
Of course, one expects that PLE still holds on the logarithmically larger
length-scale $L_*(t)$.
We shall consider that below. We caution that the above are pure PDE results,
and do not necessarily
describe statistically realizable situations. As already discussed in Section
4.2, it is not possible
that $J_0=J_1=0$ but $J_p\neq 0,\infty$ for some $p>1$ is consistent with
realizability.
Thus, only the $p=0,1$ cases of the above theorems are relevant to solutions
arising from the
Kraichnan model. If realizable initial data have a long-range power
$\alpha>d+\gamma$ and if $J_0=0$,
then it must be that $J_1\neq 0$, so the second theorem with $p=1$ applies and
the appropriately
scaled solution converges to $\Phi_1$. This agrees with the physical picture
presented in Section 3.

It would be interesting to prove spectral analogues of these theorems. We shall
not attempt
to do so here. However, we note that the conditions of the present theorems can
be implied
by spectral conditions on initial data. In fact, an initial condition with
power-law spectrum
$\sim A k^{\alpha-1}$ for an exponent $\alpha=d+\nu,\,\,\nu >\gamma$ has
spatial decay $r^{-\alpha}$
at large $r$ (except when $\nu=2j$ an even integer, when the decay may be even
faster). Thus,
$J_p\neq\infty$ for $p=0,1$. However, $J_0=0$, because the Corrsin invariant is
proportional
to the coefficient of the $k^{d-1}$ term of the low-wavenumber spectrum, which
is assumed to vanish.
Thus, the conditions of the second theorem apply, for $J_1\neq 0$. In that
case, the long-time
limit for the scaled scalar spectrum $F(\kappa,\tau(t)):=
E(\kappa/L(t),t)/(\vartheta^2(t)L(t))$
is presumably $c_1 F_1(\kappa)$ with $c_1= J_1/\widetilde{J}_1$. Although the
initial spectrum $E(k,t_0)$
had the low-wavenumber form $\sim A k^{\alpha-1}$ for $k$ somewhat smaller than
$L^{-1}(t_0)$,
at sufficiently long times the spectrum is $\sim c A(\nu,d)L^\zeta(t) k^{d+1}$
for wavenumbers
$k$ somewhat smaller than $L^{-1}(t)$, with the same constant $A(\nu,d)$ as in
(\ref{3.58}).

\noindent {\it (4.4) Initial Data with Slow Decay \& No Finite Invariant}

\noindent The last case to consider is that all of the invariants
$J_\ell,\,\,\ell=0,1,2,...$
of the initial data $\Gamma(x,0)$ are either zero or infinite. It is important
to appreciate
that this is the case for the nonexceptional scaling solutions $\Phi$
themselves. In fact,
for $p\gamma<\nu<(p+1)\gamma$, formula (\ref{3.38b}) implies that $\Phi$ has
the invariants
$\tilde{J}_\ell=0,\,\,\ell=0,...,p-1$ and $\tilde{J}_\ell=\pm \infty$ for all
$\ell\geq p.$
Because these are conserved by the dynamics, only initial data with the same
pattern of invariants can exhibit dynamical scaling at large times, governed by
a
nonexceptional $\Phi$. From this point of view the failure of PLE that was
discussed
in the last section is not so surprising. To observe a long-range power-law in
the
scaling limit, the initial data must have the same invariants as the final
scaling solution.

We prove now some theorems which establish convergence to the nonexceptional
scaling
solutions. We consider first the easiest case, when $-d<\nu<0$. We show the
following:
{\it Suppose that $\Theta(0)\in L^\infty$ and that with $0<\alpha<d$}
\be \Theta(r,0) \sim A r^{-\alpha},\,\,r\gg L_0. \lb{N1} \ee
{\it Then, if $\Gamma_{(\alpha)}(x,\tau)$ is the solution rescaled
corresponding to the parameter $\alpha,$
\be \lim_{\tau\rightarrow\infty}\Gamma_{(\alpha)}(x,\tau) = c_\alpha
\Phi_{(\alpha)}(x) \lb{N2} \ee
uniformly on compact sets. The constant $c_\alpha=A/\tilde{A}$ where
$\tilde{A}= \gamma^{\alpha/\gamma}
{{\Gamma\left({{d}\over{\gamma}}\right)}
\over{\Gamma\left(-{{\nu}\over{\gamma}}\right)}}$ is the
constant prefactor in the asymptotic power-law (\ref{3.26}) for the scaling
solution $\Phi_{(\alpha)}(\rho)
=\Phi\left({{\alpha}\over{\gamma}},{{d}\over{\gamma}};
-{{\rho^\gamma}\over{\gamma}}\right)$. Thus,
spatial PLE holds}. This seems to be the optimal result that could be expected.
The proof depends upon the observation that $\Gamma(x,0)$ may be written in the
form
\be \Gamma(x,0)= A(x,0)\Phi_{(\alpha)}(x) \lb{N3} \ee
where $A(0)\in L^\infty$ and $\lim_{x\rightarrow\infty}A(x,0)=c_\alpha
<\infty$. In fact, it follows
from the integral representation (\ref{3.41}) that $\Phi_{(\alpha)}(x)$ has no
zeros and decreases
monotonically from the value $1$ at $x=0$. This, along with asymptotic
power-law behavior
which matches that in the initial data, gives the existence of a multiplicative
perturbation
$A(x,0)$ with the required properties.

The proof uses the stochastic representation by the diffusion with drift $a(x)$
given by (\ref{C5b})
and diffusion $b(x)$ by (\ref{C5c}). By (\ref{3.41})
\be {{\Phi'_{(\alpha)}(x)}\over{\Phi_{(\alpha)}(x)}}=
    -{{\int_0^1 u e^{-ux} u^{a-1}(1-u)^{c-a-1} du}\over{\int_0^1 e^{-ux}
u^{a-1}(1-u)^{c-a-1} du}}
    \in (-1,0). \lb{N4} \ee
Furthermore, ${{\Phi'_{(\alpha)}(x)}\over{\Phi_{(\alpha)}(x)}}\sim
-{{a}\over{x}}$ for $x\gg 1$ by (\ref{3.25}).
Thus, we see that $|a(x)-c|< x$ and that for $x\gg 1$
\be a(x)\sim c-2a + x +O(x^{-1}) \lb{N5} \ee
Therefore, the drift is nonsingular except at the right boundary $x=\infty$ of
the interval,
which is infinitely attractive. The left boundary $x=0$ is repulsive. For any
$x,b$ in $(0,\infty)$
we have $\psi(x;b)= C' {{x^{-c}e^{-x}}\over{\Phi^2_{(\alpha)}(x)}}$ by
(\ref{Cx4}). Because $\psi(x;b)\sim
C' x^{-c}$ for $x\rightarrow 0+$ and $c>1$, $L_1^-=\int_0^b
\psi(x;b)\,dx=+\infty$ and the process
$X_{x,\tau}$ never reaches $0$ a.s.  However, $\psi(x;b)\sim C''
x^{2a-c}e^{-x}$ as $x\rightarrow
+\infty$, and thus $L_1^+= \int_b^\infty \psi(x;b)\,dx<+\infty$. In that case,
\cite{GS},
Section 16, Theorem 1 implies that
\be \lim_{\tau\rightarrow\infty} X_{x,\tau}=+\infty\,\,\,\,a.s. \lb{N6} \ee
Since $A(0)\in L^\infty$ and $\lim_{x\rightarrow\infty}A(x,0)=c_\alpha,$ we may
apply Lebesgue's theorem
to conclude that
\be \lim_{\tau\rightarrow \infty}A(x,\tau)
                =\lim_{\tau\rightarrow
\infty}E\left[A(X_{x,\tau},0)\right]=c_\alpha. \lb{N7} \ee
The proof is complete.

We now consider the case $\nu>0,\,\nu\neq\ell\gamma,\,\,\ell=0,1,2,...$. This
case is more
difficult because the Kummer functions $\Phi(a,c;-x)$ with $a>c$ have positive,
real zeros.
This is, of course, required by the vanishing of a certain number of the
$J$-invariants.
However, it presents a difficulty for the analysis. We prove the following:
{\it Suppose
that the initial data $\Theta(r,0)$ is a bounded perturbation of
$\Phi_{(\alpha)}$ for
$\alpha>d,\,\,\alpha\neq d+\ell\gamma,\,\,\ell=0,1,2,..$ in the sense that
\be \Gamma(\rho,0)= A(\rho,0)\Phi_{(\alpha)}(\rho), \lb{N8} \ee
with $A(0)\in L^\infty$ and that $\lim_{\rho\rightarrow\infty}A(\rho,0)=c_p
<\infty$. Suppose
also that $J_0=J_1=\cdots=J_{p-1}=0$. Then,
\be \lim_{\tau\rightarrow\infty}\Gamma_{(\alpha)}(x,\tau)= c_p
\Phi_{(\alpha)}(x) \lb{N9} \ee
uniformly on compact sets.} The conditions on $A(0)$ imply that
$\Gamma(\rho,0)\sim
A\rho^{-\alpha}$ for $\rho\gg L_0$, with $A=c_p\tilde{A}$ as before. Thus,
spatial PLE holds
for this class of initial data. However, the result is not optimal, because the
conditions
also imply that $\Gamma(\rho,0)$ has zeros at precisely the locations of the
zeros of
$\Phi_{(\alpha)}(\rho)$. The condition on vanishing invariants---which is
necessary for the
convergence result to be valid---implies the existence of such zeros, but not
at precisely
the location of the zeros of $\Phi_{(\alpha)}$. There will be more general
initial data
in the domain of attraction of the scaling solution, but there is then a
difficult problem
in accounting for the motion of the zeros in time. We leave that to future
work.

The proof of the above result uses again the stochastic representation. We note
that the
Kummer function $\Phi(a,c;-z)$ is an entire function in the complex $z$-plane
and, for
real $a,c$ has a finite number $Z$ of zeros, all simple (\cite{Erd}, Section
16). It then follows
from Weierstrass' product formula for entire functions that $\Phi(a,c;-z)=
e^{g(z)}\prod_{k=1}^Z
(z-z_k)$ where $g(z)$ is an entire function and the product is over the $Z$
complex zeros.
When $a,c>0$ the real zeros of $\Phi(a,c;-x)$ are all positive and there are
precisely $p$
such positive, real zeros $x_1,...,x_p$ when $c+p<a\leq c+p+1$ (\cite{Erd},
Section 16).  Thus,
\be {{\Phi_{(\alpha)}'(x)}\over{\Phi_{(\alpha)}(x)}}= \sum_{k=1}^p
{{1}\over{x-x_k}} + h(x) \lb{N10} \ee
where $h(x)$ is a $C^\infty$ function on the real line. For $x\gg 1$,
${{\Phi_{(\alpha)}'(x)}
\over{\Phi_{(\alpha)}(x)}}\sim -{{a}\over{x}}$ because of the asymptotic
power-law form of $\Phi_{(\alpha)}.$
Thus, we see that the drift field is
\be a(x)= c + x  + \sum_{k=1}^p {{2x}\over{x-x_k}} + 2x h(x) \lb{N11} \ee
with $p$ repulsive singular points at the zeros of $\Phi_{(\alpha)}(x)$ and, as
before,
\be a(x) \sim c-2a +x + O(x^{-1}) \lb{N12} \ee
for $x\gg 1$. Thus, the process decomposes into $p+1$ simple pieces and the
regular set of points
consists of the disjoint intervals $I_k=(x_k,x_{k+1})$, with
$x_0=0,x_p=\infty$. For each of the
first $p$ intervals, $L_1^-=L_1^+=+\infty$ as before. The process is ergodic on
these intervals
with invariant measure $P_k(x)= w_k P_{(\alpha)}(x)$ for $P_{(\alpha)}(x)=
x^{c-1}e^x \Phi_{(\alpha)}^2(x)$
and $w_k=\left[\int_{I_k} P_{(\alpha)}(x)\,dx\right]^{-1}$. However, on the
final interval
$L_1^-=+\infty$ but $L_1^{+}<+\infty$. Thus,
$\lim_{\tau\rightarrow\infty}X_{x,\tau}=+\infty$
a.s. when $x\in I_{p}$. Therefore, we have that
\be \lim_{\tau\rightarrow\infty}E\left[A(X_{x,\tau},0)\right]
                                       = w_k\int_{I_k}
A(y,0)P_{(\alpha)}(y)\,dy:= c_k \lb{N13} \ee
when $x\in I_k,\,k=0,...,p-1$ but that
\be \lim_{\tau\rightarrow\infty}E\left[A(X_{x,\tau},0)\right]= A(+\infty,0)=c_p
\lb{N14} \ee
when $x\in I_p$. We must show that $c_k=c_p$ for $k=0,1,...,p-1$. Using the
above convergence
result and the vanishing of the first $p$ invariants for the initial data, one
can show as before
that the $c$-coefficients satisfy
\be \sum_{k=0}^p c_k \int_{I_k} L_\ell^{c-1}(x)\Phi_{(\alpha)}(x)x^{c-1}\,dx=0
\lb{N15} \ee
for $\ell=0,...,p-1$. This can be written as an equation for the vector ${\bf
c}=(c_0,...,c_{p-1})^\top$
of the form ${\bf Mc}={\bf d}$ with
\be M_{\ell,k}= \int_{I_k} L_\ell^{c-1}(x)\Phi_{(\alpha)}(x)x^{c-1}\,dx
\lb{N16} \ee
for $\ell,k=0,...,p-1$ and
\be d_\ell = -c_p\int_{I_p} L_\ell^{c-1}(x)\Phi_{(\alpha)}(x)x^{c-1}\,dx
\lb{N17} \ee
for $\ell=0,...,p-1$. Because the first $p$ invariants vanish also for
$\Phi_{(\alpha)}(x)$,
one solution is $c_0=\cdots=c_{p-1}=c_p$. In fact, the matrix ${\bf M}$ is
non-singular,
because the same argument as before shows that a polynomial $p(x)$ of degree
$p-1$ which satisfies
\be \int_{I_k} p(x)\Phi_{(\alpha)}(x)x^{c-1}\,dx=0 \lb{N18} \ee
for $k=0,...,p-1$ must vanish identically, $p(x)\equiv 0$. Thus, the solution
is unique, as required.

The previous results again support the physical picture proposed in Section 3.
As before,
the theorems of this section do not describe a statistically realizable
situation when
$\alpha>d+\gamma$, although, as pure PDE results, they are valid. When the
initial data
has a long-range power decay with exponent $\alpha<d+\gamma$, then the theorems
for $p=0,1$ do apply.
In that case, two possibilities occur depending upon the value of $J_0$. The
additional
conserved quantity $J_1$ plays no role because, for $\alpha<d+\gamma$,
$J_1=\infty$ always.
If $0<J_0<\infty$, then the long-time behavior is that associated to $\Phi_0$,
as shown
in the previous section. However, if $J_0=0$ or $\infty$ (the latter will
always hold
for $\alpha<d$), then the theorems of this section imply that the long-time
behavior is
that associated to the realizable scaling solution $\Phi_{(\alpha)}$ for the
same $\alpha$.
This is the circumstance in which spatial PLE holds, in the conventional sense.

\noindent {\it (4.5) View on a Larger Length-Scale}

\noindent Finally, we will discuss briefly the scaling by $L_*(t)$
logarithmically bigger than $L(t)$.
Our analysis will be only heuristic and no proofs given. The solution to the
equation (\ref{3.6})
is $L(t)=\gamma D_1(t-t_0^*)^{1/\gamma}$ for a virtual time origin $t_0^*$
related to the initial data
$L_0$. Let us define
\be  L_*(t)=\gamma D_1\left[(t-t_0^*)\log(t-t_0^*)\right]^{1/\gamma}. \lb{X0}
\ee
We now study rescaled solutions
\be  \Theta(r,t) = \vartheta^{2}(t)\Gamma_*(r/L_*(t),\tau(t)), \label{X1} \ee
where $\vartheta(t),\tau(t)$ are as before. In particular, $\tau(t): = \log
(t-t_0^*)=
L_*^\gamma(t)/L^\gamma(t)$. The function $\Gamma_*$ solves:
\be
 \gamma {{\partial
\Gamma_*}\over{\partial\tau}}(\rho,\tau)={{1}\over{\tau}}{{1}\over{\rho^{d-1}}}
{{\partial}\over{\partial\rho}}\left[ \rho^{d+\zeta -1} {{\partial
\Gamma_*}\over{\partial\rho}}
 (\rho,\tau)\right] + \left(1+{{1}\over{\tau}}\right)\rho{{\partial
\Gamma_*}\over{\partial\rho}}(\rho,\tau)
 +\alpha \Gamma_*(\rho,\tau), \label{X3}
\ee
where $\rho = r/L_*(t)$. If we make the change of variables $x =
\gamma^{-1}\rho^{\gamma}$, we obtain
\be {{\partial \Gamma_*}\over{\partial\tau}}(x,\tau)
     =  x{{\partial\Gamma_*}\over{\partial x}}(x,\tau) +a \Gamma_*(x,\tau)
        +{{1}\over{\tau}}\left[ x {{\partial^2\Gamma_*}\over{\partial
x^2}}(x,\tau)
        + (c+x){{\partial\Gamma_*}\over{\partial x}}(x,\tau)\right]x,
\label{X4} \ee
with $a=\alpha/\gamma,c=d/\gamma$.

If the initial datum has the power-law form $\Gamma(x,0)\sim A x^{-a}$ at large
$x$, then we expect
that $\Gamma_*(x,\tau)$ converges to that same power pointwise in $x$. This
motivates us to define
$A_*(x,\tau)$ via
\be \Gamma_*(x,\tau)= x^{-a} A_*(x,\tau). \lb{X5} \ee
Thus, we want to show that $\lim_{\tau\rightarrow\infty}A_*(x,\tau)=
A(+\infty,0)$. Substitution of
(\ref{X5}) into (\ref{X4}) yields the equation
\be {{\partial A_*}\over{\partial\tau}}(x,\tau)
     =  x{{\partial\Gamma_*}\over{\partial x}}(x,\tau)
        +{{1}\over{\tau}}\left[ x {{\partial^2\Gamma_*}\over{\partial
x^2}}(x,\tau)
        + (c-2a+x){{\partial\Gamma_*}\over{\partial x}}(x,\tau)
        + {{a(a-c+1-x)}\over{x}}A_*(x,\tau)\right]. \label{X6} \ee
Heuristically, we may expect that, at long times, the terms in the bracket can
be neglected
because of the factor $1/\tau$. In that case, the solution is governed by just
the hyperbolic
equation ${{\partial A_*}\over{\partial\tau}}(x,\tau)=
x{{\partial\Gamma_*}\over{\partial x}}(x,\tau)$.
This is solved by the method of characteristics, as $A_*(x,\tau)=A(e^\tau
x,0)$. If this crude
approximation is valid, then
$\lim_{\tau\rightarrow\infty}A_*(x,\tau)=A(+\infty,0)$, as required.

A more refined estimate can be made by using a stochastic representation
corresponding to
the diffusion process with time-dependent drift
\be a(x,\tau)= x + {{c-2a+x}\over{\tau}} \lb{X7} \ee
and diffusion
\be b(x,\tau)=\sigma^2(x,\tau)= {{2x}\over{\tau}}. \lb{X8} \ee
If $X_{x,\tau}$ is the process started at $x$ at time $\tau_0$, the solution of
(\ref{X6})
has the representation
\be A_*(x,\tau)=E\left[ \exp\left(\int_{\tau_0}^\tau
c(X_{x,\sigma},\sigma)d\sigma\right)
                        A(X_{x,\tau},0)\right], \lb{X9} \ee
where
\be c(x,\tau)= {{a(a-c+1-x)}\over{x\cdot\tau}}. \lb{X10} \ee
See \cite{GS}, Section 11, Theorem 3. Because the drift $a(x,\tau)\sim x$ at
large $x,\tau$ and,
furthermore, $\sigma^2(x)/x^2\rightarrow 0$ as $x\rightarrow\infty$ or
$\tau\rightarrow\infty$, one
expects that $X_{x,\tau}\sim e^\tau$ as $\tau\rightarrow\infty$ a.s. See
\cite{GS}, Section 17,
Corollary 2. The contribution to the integral of $c(x,\sigma)$ in the exponent
for $\sigma {\,}^<_\sim
\tau$ may be estimated by $c(X_{x,\sigma},\sigma)\approx -a/\sigma$ when
$X_{x,\sigma}$ is large,
as it is with high probability when $\sigma {\,}^<_\sim\tau$. This gives a term
$\approx \exp(-a\log\tau)= \tau^{-a}$, which vanishes for large $\tau$. On the
other hand,
the contribution from $\sigma{\,}^>_\sim\tau_0$ is of the order of
$\exp\left(O(1)/\tau\right)
\sim 1$ for large $\tau$. Thus, we expect again that
$\lim_{\tau\rightarrow\infty}A_*(x,\tau)=
A(+\infty,0).$

\newpage

\section{Conclusions}

In this paper we have studied in detail the 2-point correlation function in the
Kraichnan model, which, for homogeneous and isotropic conditions, is governed
by
the equation (\ref{1.3}). Our main technical results are as follows:
\begin{itemize}
\item[(i)] We have found all the possible self-similar decay solutions,
parameterized by
          space dimension $d$, velocity field roughness exponent
$\zeta=2-\gamma$, and
          a scaling exponent $\alpha>0$, in terms of Kummer confluent
hypergeometric functions.
\item[(ii)] We have shown that these solutions are statistically realizable
           for all $\alpha$ in the interval $(0,d+\gamma]$ and for no $\alpha$
in the
           complement $(d+\gamma,\infty)$.
\item[(iii)] We have exhibited an infinite sequence of invariants
$J_\ell,\,\,\ell=0,1,2,...$
            of the equation (\ref{1.3}), defined by the formula (\ref{C5n}).
$J_0$ coincides
            with the well-known Corrsin invariant. At least one of
            $J_0,J_1$ must be nonzero (possibly infinite) for
            realizable initial data.
\item[(iv)] In terms of these invariants, we have identified
            initial data in domains of attraction of the scaling
            solutions found in (i). Our theorems relevant to the
            Kraichnan model covered the following cases:
$$ \,\,\,\,\,\,\,\,\,\,\,\,
   \begin{array}{ll}
   J_0=J_1=\infty &
(\Gamma_{(\alpha)}(\tau)\rightarrow\Phi_{(\alpha)},\,\,0<\alpha<d) \cr
   J_0=0,J_1=\infty &
(\Gamma_{(\alpha)}(\tau)\rightarrow\Phi_{(\alpha)},\,\,d<\alpha<d+\gamma) \cr
   0<J_0<\infty,J_1\,\,{\rm arbitrary} &
(\Gamma_{(\alpha)}(\tau)\rightarrow\Phi_0,\,\,\alpha=d) \cr
   J_0=0,0<J_1<\infty &
(\Gamma_{(\alpha)}(\tau)\rightarrow\Phi_1,\,\,\alpha=d+\gamma)
   \end{array} $$
            The first two results apply for initial data with a
            power-law decay $\sim r^{-\alpha}$ for $\alpha<d+\gamma$,
            whereas the last two include (among other possibilities)
            initial power-laws with $\alpha>d+\gamma$.
\end{itemize}
There are several questions left open in this work that deserve to be
addressed. It would be
worthwhile to establish convergence for more general initial data with slow
decay and no finite
invariants, allowing for a non-coincidence of its zeros with those of the final
scaling solution.
It would be interesting as well to extend the convergence results to the
spectral domain and to establish
convergence to a power-law solution on the logarithmically larger scale.
Perhaps the most interesting
issues not treated in the present work are to consider scalar decay in the
Kraichnan model with initial
anisotropy and/or inhomogeneity and to consider the self-similarity (or not) of
the higher-order
$N$-point scalar correlations with $N>2$. The latter is relevant for realistic
decay problems,
where no closure of the moment hierarchy is found. In fact, von K\'{a}rm\'{a}n
and Howarth
in their classic analysis \cite{vKH} made a self-similarity hypothesis for the
3rd-order as well
as the 2nd-order velocity correlations, in order to derive a scaling
equation.

Perhaps the most novel result in this work is the discovery of the infinite
sequence of
invariants $J_\ell,\,\,\ell=0,1,2,...$ in the Kraichnan model. It would be most
interesting
to know whether there are analogues of such invariants for true passive
scalars, besides the
Corrsin invariant $K=J_0$, or, for that matter, for Navier-Stokes turbulence.
In the latter
case, the integral invariant $C=\int_{{\Bbb R}^d} d^d\br\,\,B_{LL}(r)$ found by
Saffman \cite{Saff}
is analogous to the Corrsin invariant, while the integral $\Lambda=\int_{{\Bbb
R}^d} d^d\br\,\,r^2
B_{LL}(r)$ found by Loitsyanskii \cite{Loit} is rather analogous to our next
invariant $J_1$.
However, it is well-known that $\Lambda(t)$ is actually time-dependent, due to
the effects
of spatial long-range correlations \cite{PR}. The connection of the
``generalized invariants''
introduced here with the ``zero modes'' studied in \cite{BGK} is very
intriguing,
especially as the latter exist also for anisotropic statistics and,
plausibly, for higher-order $N$-point correlators.

Our results confirm in the Kraichnan model the ``two-scale'' decay picture
earlier found
for Burgers equation in \cite{GSAFT}. The fact that there are no realizable
scaling solutions for $\alpha>d+\gamma$, implies that initial data with such a
$r^{-\alpha}$
decay at large distance cannot exhibit a self-similar decay with the same
power-law behavior
on the scale of the integral length $L(t)$. Instead, we have shown that a wide
class of such
initial data converge on that length-scale to the self-similar decay solution
$\Theta_1(r,t)$
corresponding to $\alpha=d+\gamma$. A heuristic argument supports the
conclusion that the initial
power-law decay will be preserved on a length-scale $L_*(t),$ which is
logarithmically bigger than $L(t)$.
The fact that the same phenomenon occurs for two such very different models as
Burgers equation
and Kraichnan's passive scalar model argues for its generality. In fact, the
physics behind
it appears to be just the phenomenon of ``backtransfer'', which closure results
indicate
is a very general feature of turbulent decay. The specific, simplifying
features of the models
only enter in calculating the rates of this backtransfer.

We therefore expect that the results of this work will have relevance for real
passive scalars
and other turbulent decay problems. It is thus of some interest to specialize
our results
to that case of the Kraichnan model which most closely mimicks the true passive
scalar, namely,
$d=3$ and $\zeta={{4}\over{3}}$. In terms of the scalar low-wavenumber spectral
exponent $n=\alpha-1$,
there is a self-similar decay with PLE if initially $-1<n<{{8}\over{3}}$.
However, if ${{8}\over{3}}<n<4$
then these traditional expectations are violated. In the whole range
$n>{{8}\over{3}}$
the scalar energy decay law is $E(t)\sim (t-t_0)^{-11/2}$, as traditionally
expected
only for the lowest value $n={{8}\over{3}}$. We have learned of unpublished
work of
D. Thomson who finds exactly the same behavior in a simple model of a mandoline
source
\cite{Thom1,Thom2}. This leads us to believe that the results presented here
have some validity
outside the white-noise model. In fact, on this basis, we have presented a
phenomenological
extension of this picture to decay of 3D incompressible fluid turbulence
\cite{EyTh}.

Although the physics will be different, similar phenomena may occur in other
nonequilibrium
decay processes. During the preparation of this paper we became aware of a
``two-scale'' picture
for phase-ordering by Cahn-Hilliard dynamics with power-law correlated (or
fractally clustered) initial
data \cite{CMS}. From a renormalization group (RG) point of view, our results
correspond to a case
where some of the ``fixed points'', i.e. the scaling solutions, are unphysical.
A careful
study of this example using general methodologies such as RG may help to
illuminate
subtleties in their application.

\vspace{.1in}

\noindent {\Large\bf Acknowledgements}. G. E. wishes to acknowledge the Isaac
Newton Institute of
Mathematical Sciences for its hospitality during his stay for the Programme on
Turbulence,
Jan.4-July 2, 1999, when part of this work was done. Also, he wishes to thank
M. Chertkov,
U. Frisch, N. Goldenfeld, and D. J. Thomson for helpful discussions.
J. X. was partially supported by NSF grant DMS-9625680
and would like to thank J. Watkins for useful conversations.

\newpage

\noindent {\bf APPENDIX:}

\noindent{\bf SELF-SIMILAR SCALAR SPECTRA FOR A BROWNIAN VELOCITY FIELD}

\vspace{.2in}

\noindent In general, it is not possible to evaluate explicitly the Fourier
transforms of the
scaling functions, $\widehat{\Phi}(\kappa),$ not even in terms of standard
special functions.
This is not possible even for the equipartition case $\alpha=d$, in which case
the Fourier transform
is the spherically symmetric L\'{e}vy stable distribution with parameter
$\gamma$ in $d$ dimensions.
However, it is known that the stable distributions are calculable explicitly
for the parameter
value $\gamma=1$, in which case they are the $d$-{\it dimensional Cauchy
distributions}.
Since then also $\zeta=1$, this case corresponds to an advecting velocity field
with
H\"{o}lder exponent $H={{1}\over{2}}$, or a Brownian-type random field.

In fact, the Cauchy distribution of parameter $\beta$ in $d$-dimensions is
defined by the Fourier transform
\be G(\kappa;\beta)={{1}\over{(2\pi)^{d/2}\kappa^{(d-2)/2}}}\int_0^\infty
                    e^{-\beta\rho}J_{(d-2)/2}(\kappa\rho)\,\rho^{d/2}\,d\rho
\lb{A1} \ee
with the result
\be G(\kappa;\beta)=
{{2}\over{(4\pi)^{d/2}}}
                     {{\Gamma(d)}\over{\Gamma\left({{d}\over{2}}\right)}}\cdot
                     {{\beta}\over{(\beta^2+\kappa^2)^{{{1}\over{2}}(d+1)}}}.
\lb{A2} \ee
This follows from \cite{Erd},7.7.3(16) with the parameter choices
$\gamma=\beta,\alpha=\kappa,
\mu={{1}\over{2}}(d-2),\rho={{1}\over{2}}(d+2)$ there, because the
hypergeometric series then
simplifies to a binomial series for the function
$\left(1+{{\kappa^2}\over{\beta^2}}\right)^{
-{{1}\over{2}}(d+1)}$. We shall exploit this fact to evaluate the Fourier
transforms
$\widehat{\Phi}(\kappa)$ for $\gamma=1$, at least for all of the exceptional
cases $\alpha=d+\ell,
\,\ell=0,1,2,...$. We may use $G(\kappa;\beta)$ as a generating function in
those cases, because
\be \widehat{\Phi}_\ell(\kappa)=
\left.{{1}\over{(d)_\ell\beta^{d-1}}}{{\partial^\ell}\over{\partial\beta^\ell}}
    \left[\beta^{d+\ell-1}G(\kappa;\beta)\right]\right|_{\beta=1}, \lb{A3} \ee
as a consequence of (\ref{3.29}).

The results obtained by use of this formula are, explicitly, for $\ell=0$,
\be \widehat{\Phi}_0(\kappa)=
{{2}\over{(4\pi)^{d/2}}}{{\Gamma(d)}
                             \over{\Gamma\left({{d}\over{2}}\right)}}\cdot
                            {{1}\over{(1+\kappa^2)^{{{1}\over{2}}(d+1)}}},
\lb{A4} \ee
for $\ell=1$,
\be \widehat{\Phi}_1(\kappa)=
{{2}\over{(4\pi)^{d/2}}}{{\Gamma(d)}
\over{\Gamma\left({{d}\over{2}}\right)}}\cdot
{{(d+1)\kappa^2}\over{d(1+\kappa^2)^{{{1}\over{2}}(d+3)}}}, \lb{A5} \ee
for $\ell=2$,
\be \widehat{\Phi}_2(\kappa)=
{{2}\over{(4\pi)^{d/2}}}{{\Gamma(d)}
        \over{\Gamma\left({{d}\over{2}}\right)}}\cdot
        {{(d+2)\kappa^4-\kappa^2}\over{d(1+\kappa^2)^{{{1}\over{2}}(d+5)}}},
\lb{A6} \ee
for $\ell=3$,
\be \widehat{\Phi}_3(\kappa)=
{{2}\over{(4\pi)^{d/2}}}{{\Gamma(d)}
\over{\Gamma\left({{d}\over{2}}\right)}}\cdot
{{(d+3)[(d+2)\kappa^6-3\kappa^4]}
\over{d(d+2)(1+\kappa^2)^{{{1}\over{2}}(d+7)}}}, \lb{A7} \ee
and for $\ell=4$,
\be \widehat{\Phi}_4(\kappa)=
{{2}\over{(4\pi)^{d/2}}}{{\Gamma(d)}
\over{\Gamma\left({{d}\over{2}}\right)}}\cdot
{{(d+4)(d+2)\kappa^8-6(d+4)\kappa^6+3\kappa^4}
\over{d(d+2)(1+\kappa^2)^{{{1}\over{2}}(d+9)}}}. \lb{A8} \ee
Of course, these transforms and those for all $\ell$ are analytic at $\kappa=0$
with radius of convergence
of the power series equal to 1. It may be easily checked that the coefficients
are equal to the
$B_j$ given by (\ref{3.37c}). In particular, there is no singular contribution
at $\kappa=0$.

These explicit results illustrate several of the general conclusions reached in
the text.
For example, we see that $\widehat{\Phi}_0(\kappa),\widehat{\Phi}_1(\kappa)$
are everywhere
nonnegative and that
$\widehat{\Phi}_1(\kappa)=-{{1}\over{d}}\kappa{{\partial\widehat{\Phi}_0}
\over{\partial\kappa}}(\kappa).$ Furthermore, in agreement with the result that
realizability fails
whenever $\nu>\gamma$, we see that the Fourier transforms for $\ell=2,3,4$ are
{\it not} everywhere
positive. In the cases $\ell=2,3$ the coefficient of the dominant
low-wavenumber power is
negative, so realizability fails for the lowest wavenumber range. However, for
$\ell=4$,
realizability fails despite the coefficient of the dominant low-wavenumber
power being positive.
In fact, the polynomial
\be P(\kappa^2)=(d+4)(d+2)\kappa^8-6(d+4)\kappa^6+3\kappa^4 \lb{A9} \ee
has two positive roots $\kappa_\pm^2=
{{3}\over{d+2}}\left[1\pm\sqrt{{{2(d+5)}\over{3(d+4)}}}\,\,\right]$
and is negative in the interval $(\kappa_-^2,\kappa_+^2)$.

\end{document}